\documentclass[aps,prc,eqsecnum,amsmath,onecolumn,notitlepage,nofootinbib,superscriptaddress,10pt]{revtex4-2}

\usepackage{lineno}
\usepackage{graphicx}
\usepackage{bm}
\usepackage{natbib}
\usepackage{amsmath}
\usepackage{epsfig}
\usepackage{url}
\usepackage{accents}
\allowdisplaybreaks
\usepackage{xcolor}
\usepackage{comment}

\def\ang#1{\langle#1\rangle }
\def\w#1#2{W^{[#1#2]}}
\def\W#1#2#3{W^{[#1#2#3]}}

\def\Lang#1#2{L^{[#1]}_{\;[#2]}}

\def\LLang#1#2#3{L^{[#1]}_{\;[#2 #3]}}
\def\vq{{\bm q}}
\def\vx{{\bm x}}
\def\vy{{\bm y}}
\def\tvq{{\tilde{\bm q}}}
\def\vt{{\bm t}}
\def\csp{w^\prime}
\def\geta{\gamma_\eta}

\def\vpi{{\bm \delta\pi}}

\begin{document}

\title{Non-Gaussian hydrodynamic fluctuations in an expanding relativistic fluid}

 \author{G\"{o}k\c{c}e Ba\c{s}ar}
 \email{gbasar@unc.edu}
 \affiliation{Department of Physics and Astronomy, University of North Carolina,
Chapel Hill, North Carolina 27599, USA}

\author{Shuo Song}
\email{shuosong@live.unc.edu}
\affiliation{Department of Physics and Astronomy, University of North Carolina,
Chapel Hill, North Carolina 27599, USA}

\begin{abstract}
We consider non-equilibrium evolution of non-Gaussian fluctuations in a hydrodynamic system undergoing a boost-invariant expansion described by Bjorken flow. We derive the evolution equations for two‑ and three‑point velocity correlators using the effective field‑theory framework and present analytical solutions for them. We show that the average
Landau frame is better suited for studying non-Gaussian fluctuations of velocity when relativistic effects are important. In the Bjorken background, the average Landau frame corresponds to the density frame.  We demonstrate that the three‑point correlators depend nonlinearly on the non‑equilibrium dynamics of the two‑point functions, and exhibit non-trivial effects such as memory. The importance of these effects in the context of the search for the QCD critical point via fluctuations is discussed. 
\end{abstract}

\maketitle

\section{Introduction}

The theory of hydrodynamics constitutes a universal long‑wavelength, low‑frequency description of many‑body systems near thermal equilibrium, whose applicability spans sub‑atomic to astrophysical scales. In particular, relativistic hydrodynamics, as studied in this paper, has been shown to successfully describe the physics of the quark‑gluon plasma created in heavy‑ion collision experiments \cite{Romatschke:2017ejr,Jeon:2015dfa}.

In general, hydrodynamics includes dissipation, which necessitates the existence of thermal fluctuations via the fluctuation–dissipation theorem. Over the past decade, the theory of thermal fluctuations in relativistic hydrodynamics has attracted significant interest, largely driven by relativistic heavy‑ion collision experiments \footnote{For a review of the recent progress on relativistic hydrodynamic fluctuations see Ref. \cite{Basar:2024srd}.}. 

One of the primary objectives of the Beam Energy Scan program at the Relativistic Heavy Ion Collider is to map the QCD phase diagram and, in particular, to search for the QCD critical point \cite{Luo_2016}. A key feature of the critical point is the parametric enhancement of thermal fluctuations in its vicinity. Consequently, measurements of fluctuation observables in heavy‑ion collisions provide a promising experimental avenue to search for the critical point \cite{Stephanov:1999zu,STAR:2010vob,An:2021wof,Stephanov:2024xkn}. Most state‑of‑the‑art theoretical predictions for fluctuation observables sensitive to criticality assume thermal equilibrium \cite{Bzdak:2019pkr}. However, the droplet of quark‑gluon plasma created in heavy‑ion collisions undergoes rapid expansion and cooling, resulting in a highly dynamical medium. Because thermalization does not occur instantaneously, not all fluctuation modes have sufficient time to reach equilibrium before the system cools down to the freeze‑out temperature, where hadrons are produced. As a result, non‑equilibrium effects, such as memory, sourced by the dynamical evolution of fluctuations play a crucial role in making quantitative predictions for the  QCD critical point signatures. Capturing these effects requires a dynamical description of fluctuations in an expanding background.

The approach adopted in this paper is to derive deterministic time‑evolution equations for the moments of the probability distribution that encodes hydrodynamic fluctuations. These moments correspond to correlation functions of conserved densities, whose evolution equations resemble kinetic equations describing the propagation and damping of phonon‑like modes. This so‑called \textit{hydro‑kinetic} approach to relativistic hydrodynamic fluctuations was introduced in Ref.~\cite{akamatsu} to derive evolution equations for Gaussian fluctuations (two‑point correlation functions) in a fluid undergoing one‑dimensional, boost‑invariant expansion (Bjorken flow), and was later generalized to include a conserved baryon charge in Ref.~\cite{martinez_2019}. 

In Refs.~\cite{An:2019osr,An:2019csj}, evolution equations for Gaussian fluctuations in arbitrary fluid backgrounds were derived in a Lorentz‑covariant form, referred to as the \textit{confluent formalism}. This framework was subsequently extended to non‑Gaussian fluctuations in Ref. ~\cite{An:2020vri},  where higher‑order correlators were derived for nonlinear charge diffusion, and further developed in Ref.~\cite{An:2022jgc} to describe the non‑Gaussian dynamics of the slowest hydrodynamic mode in a general background. Most recently, the confluent formalism was extended to include non‑Gaussian fluctuations of all hydrodynamic modes in arbitrary backgrounds \cite{An:2026glk}. 

As in the works cited above, we remain in the hydrodynamic regime, where the characteristic spatial scale of inhomogeneities sourced by fluctuations is large compared to the correlation length. In this limit, fluctuations are present but parametrically small. The smallness of fluctuations is controlled by the ratio of the correlation length, $\xi$, to the wavelength $l$, of a given fluctuation mode, quantified by the dimensionless parameter $\epsilon \equiv (l/\xi)^3$ which corresponds to the inverse of the number of statistically independent fluid cells \cite{An:2022jgc,An:2020vri}. In the hydrodynamic limit, the central limit theorem implies that $\epsilon\ll1$, and fluctuations are suppressed by powers of $\epsilon$.

Parallel to these developments, significant progress has been made in formulating dissipative relativistic hydrodynamics as an effective (statistical) field theory based on an action principle \cite{Kovtun:2014hpa,Crossley:2015evo,Haehl:2018lcu,Glorioso_2017,Jain:2020hcu}.In this formalism, the effective action describes the long-lived, long distance modes of the fluid and includes all possible terms consistent with the underlying symmetries, including crucially the local Kubo-Martin-Schwinger symmetry that ensures that the fluctuation-dissipation relation is satisfied locally. As in any effective field theory, the action is organized as a derivative expansion reflecting the separation between microscopic and macroscopic scales. For a pedagogical introduction to the subject we refer the reader to Ref. \cite{liu-lecture}.
Within this EFT formalism, evolution equations for fluctuation correlators arise as truncated Schwinger–Dyson equations of the effective theory. This framework was previously applied to nonlinear diffusive systems in Ref.~\cite{Sogabe_2022} where the role of higher‑order noise terms beyond Gaussian order -- originally introduced in Ref. \cite{jain2022} --  were also discussed. 

The primary goal of this paper is to provide analytical expressions for the dynamics of non‑Gaussian hydrodynamic fluctuations in an expanding background, suitable for phenomenological applications. In particular, these results can serve as input for the maximum‑entropy freeze‑out framework \cite{Pradeep:2022eil,Karthein:2025ini,Basar:2026irk}  which converts hydrodynamic fluctuations into hadronic fluctuations which are experimentally accessible. As a first step toward this goal, we derive the evolution equations for non‑Gaussian hydrodynamic fluctuations in the Bjorken background using the EFT approach,  solve them, and highlight their essential dynamical features.

 In Section \ref{sec:fluctuations} we review the basics of hydrodynamic fluctuations and introduce the general formalism for fluctuating hydrodynamics in the EFT framework. In Sec. \ref{sec:evolution} we apply the formalism to the Bjorken background and derive the three-point functions for the energy density and velocity fluctuations. The solutions to the evolution equations are presented in Section \ref{sec:analytical} along with their universal relaxation to equilibrium. 
 We summarize our results and discuss the outlook for future developments in Section \ref{sec:conclusions}.
In Appendix \ref{sec:evolution} we review the simple example of a particle undergoing Brownian motion in the hydro-kinetic as well as EFT frameworks as a pedagogical example.

\section{Hydrodynamic Fluctuations}
\label{sec:fluctuations}

\subsection{Generalities}
\label{sec:generalities}

Thermal fluctuations can be incorporated into hydrodynamics in several complementary ways. A classical approach \cite{landau1980statistical} is to introduces Gaussian noise directly into the hydrodynamic equations, promoting them to stochastic differential equations. 
In this framework, the hydrodynamic energy momentum tensor differs from what the constitutive equations predict by a random amount,
\begin{equation}
    T^{\mu\nu}= T^{\mu\nu}_{\rm ideal}+T^{\mu\nu}_{\rm visc}+S^{\mu\nu},
\end{equation}
where the ideal and viscous contributions are given by \begin{equation}
\label{eq:constitutive}
    T^{\mu\nu}_{\rm ideal}=\varepsilon_L u^\mu u^\nu + p\Delta^{\mu\nu},\quad T^{\mu\nu}_{\rm visc}=\eta[\Delta^{\mu\sigma}\nabla_\sigma u^\nu + \Delta^{\nu\sigma}\nabla_\sigma u^\mu] -(\zeta - \frac{2}{3}\eta)\Delta^{\mu\nu}\nabla_\sigma u^\sigma\,.
\end{equation}
Here $\varepsilon_L$, $p$, $u^\mu$, denote the energy density, pressure, and fluid four‑velocity, normalized as $u^2=-1$.  The subscript $L$ stands for ``Landau frame" in which the fluid velocity is aligned with the energy flow,
\begin{equation}
    u_\mu T^{\mu\nu}=-\varepsilon_L u^\nu\,.
\end{equation}
The shear and bulk viscosities are denoted by $\eta$ and $\zeta$, and $\Delta^{\mu\nu}=g^{\mu\nu}+u^\mu u^\nu$ projects orthogonally to the fluid velocity as usual. The stochastic contribution to the energy momentum tensor, $S^{\mu\nu}$, is local and sampled from a Gaussian distribution
\begin{equation}
    \langle S^{\mu\nu}(x)S^{\alpha\beta}(x^\prime) \rangle = T\left[2\eta \Delta^{\mu(\alpha}\Delta^{\beta)\nu}  + \left(\zeta - \frac{2}{3}\eta\right)\Delta^{\mu\nu}\Delta^{\alpha\beta}\right]\delta^{(4)}(x-x^\prime)
\end{equation}
where the average is over an ensemble of configurations. Hydrodynamic evolution in the presence of noise then follows from energy–momentum conservation, $\partial_\mu T^{\mu\nu}=0$,
 yielding stochastic equations for$\varepsilon_L$, $u^\mu$, analogous to Langevin dynamics.

An alternative and more systematic formulation is provided by the effective field theory (EFT) approach to fluctuating hydrodynamics, formulated within the Schwinger–Keldysh framework \cite{liu-lecture}. In this construction, hydrodynamics is described by an effective action constrained by the symmetries of the underlying microscopic theory, including dynamical Kubo–Martin–Schwinger (KMS) symmetry, which enforces the fluctuation–dissipation relation locally. The effective action is organized as a derivative expansion stemming from the hydrodynamic separation of scales. 

In addition to the usual hydrodynamic variables, the EFT introduces an auxiliary vector field $X_a^\mu$ such that the effective Lagrangian takes the form\cite{liu-lecture}
\begin{equation}
    \begin{split}
        \mathcal{L} =&  \left[\varepsilon_L u^\mu u^\nu + p\Delta^{\mu\nu} 
        - \eta(\Delta^{\mu\sigma}\nabla_\sigma u^\nu + \Delta^{\nu\sigma}\nabla_\sigma u^\mu) 
        -(\zeta - \frac{2}{3}\eta)\Delta^{\mu\nu}\nabla_\sigma u^\sigma \right] \nabla_\mu X_{a\nu} \\
    & + iT[2\eta \Delta^{\mu(\alpha}\Delta^{\beta)\nu}  + (\zeta - \frac{2}{3}\eta)\Delta^{\mu\nu}\Delta^{\alpha\beta}]\nabla_\mu X_{a\nu} \nabla_\alpha X_{a\beta}\,.
    \end{split}
\label{eq:Landau-Lagrangian}
\end{equation}
 To leading order in $X_a$, the equations of motion simply reduce to the classical conservation laws. Terms quadratic in $X_a$ generate Gaussian noise with coefficients fixed by dissipative transport parameters. Higher‑order noise terms can be organized into KMS‑invariant blocks \cite{jain2022}. Unlike the Gaussian noise term, the coefficients of these blocks are not universally fixed by hydrodynamics and have to be determined from the underlying microscopic theory. However, for equal‑time correlation functions—our focus here—such terms contribute only at higher order in the gradient expansion and can be neglected \cite{Sogabe_2022}.  

The central objects of interest in this paper are equal‑time correlation functions of hydrodynamic fluctuations,
\begin{equation}
    G_{A_1A_2\dots A_N}(\vx_1,\vx_2,\dots \vx_N;t)=\langle \phi_{A_1}(\vx_1,t)\phi_{A_2}(\vx_2,t)\dots \phi_{A_N}(\vx_N,t) \rangle.
\end{equation}
where the field $\phi_{A}$ collectively denotes the energy and momentum density fluctuations. These correlators describe the dynamical evolution of fluctuations on top of an average fluid background. 
Because hydrodynamics relies on a separation between microscopic and macroscopic scales, fluctuation correlators exhibit a natural hierarchy.  The spatial variations occur in a length scale, $l$, which is much greater than the microscopic scale $l_{mic}$.  This naturally gives rise to a small parameter $l_{mic}/l$, which controls hydrodynamic the gradient expansion. At the same time,  hydrodynamic fluctuations occur in a typical spatial inhomogeneity scale proportional to inverse wavelength, $1/q$, which is also much larger than $l_{mic}$. The magnitude of the fluctuations, averaged over the scale $1/q$, is suppressed by the inverse square root of the number of uncorrelated cells as a result of the central limit theorem. Since the correlation length, $\xi$, is proportional to $l_{mic}$ as well, the number of uncorrelated cells is $(q l_{mic})^3\equiv 1/\epsilon$
 \footnote{In the vicinity of a critical point $\xi\sim l$, and this hierarchy of scales breaks down and non-Gaussian fluctuations cease to be suppressed in magnitude, a hallmark of critical phenomena. }. Therefore magnitude of fluctuations is suppressed by $\sqrt{\epsilon}$.  
 This suppression naturally generates a hierarchy between the higher-point correlators such that the $N$-point function scales as\cite{An:2020vri,An:2022jgc}:
\begin{equation}
    G_{A_1,\dots,A_N} \sim \epsilon^{[N/2]}
\end{equation}
where $[N/2]$ equals $N/2$ for even $N$ and $(N+1)/2$ for odd $N$. At the same time, the \textit{connected} correlators, which are the main objects of our interest in this paper, scale as
\begin{equation}
    G^c_{A_1,\dots,A_N} \sim \epsilon^{N-1}
\end{equation}
As a result, we can systematically truncate higher-order terms from the evolution equations. After the truncation, the evolution equation for an $N-$point function only involves contributions from $k\leq N$-point functions which we can schematically express  as 
\begin{equation}
    \partial_t G_N={\cal F}_N[G_2,G_3,\dots, G_N]\,.
    \label{eq:GN_evolve}
\end{equation}
 The right-hand side is linear in $G_N$, but includes nonlinear combinations of lower-point functions and their spatial gradients. Also important is that the lower point functions appear as inhomogeneous terms for the equation for $G_N$, allowing the set of equations to be solved iteratively. 

In this work, we shall focus on the evolution equations of the three-point correlators in the background of Bjorken flow. The equations for two-point functions have first been studied in the Bjorken background in the seminal work of Ref. \cite{akamatsu} and have been generalized to fluids with conserved baryon charge by Ref. \cite{martinez_2019} and to an arbitrary background flow in Refs. \cite{An:2020vri,An:2022jgc}.

\subsection{Deterministic time evolution in noisy Lorentz frames}

The hydro-kinetic approach aims to describe fluctuations through deterministic evolution equations for correlation functions. A subtlety arises because the natural “time” direction in the local rest frame of the fluid is defined by the fluid four‑velocity ,$u^\mu$, which itself fluctuates. While this poses no difficulty for two‑point functions, it leads to ambiguities for higher‑order correlators. One common strategy is to define the time direction using the \textit{average} fluid velocity, $\bar u^\mu$, as adopted in Refs.~\cite{An:2019csj, An:2019osr, An:2022jgc}. The fluid velocity then can be written as
\begin{equation}
    u^\mu=\bar u^\mu+\delta u^\mu,
\end{equation} 
where $\delta u^\mu$ denotes fluctuations around average. The stochastic noise is purely spatial in the instantaneous local rest frame, satisfying $u_\mu S^{\mu\nu}=0$. However, when expressed in a frame defined by, $\bar u^\mu$, the noise acquires a time‑like component, 
\begin{equation}
    \bar u_\mu S^{\mu\nu}=-\delta u_\mu S^{\mu\nu}
\end{equation}
which generates time‑derivative terms of the noise in the evolution equations.  A quick way to see this is to notice that the evolution equations sketched in Eq. \eqref{eq:GN_evolve}, in essence, follow from Ward type identities
\begin{equation}
    \langle\nabla_\mu T^{\mu\nu}(\vx;t) \phi_1({\bm x_1};t)\dots\phi_k ({\bm x_1};t) \rangle=0
    \label{eq:ward}
\end{equation}
where $\phi_i$s are given in terms of hydrodynamic fields as we  discuss further in Sec. \ref{sec:SD} \footnote{The equal time limit has to be taken with care, but it is not of importance for this discussion.}. The timelike component of $S^{\mu\nu}$ therefore generates a term  $\bar u\cdot \partial (\bar u_\mu S^{\mu\nu})$ which contributes to the evolution equation as
\begin{equation}
    \bar u\cdot \nabla G_N=\langle \bar u\cdot \nabla_\mu  (\delta u_\mu S^{\mu\nu})\dots\rangle+\dots
\end{equation}
 Since the noise is $\delta$-correlated in time, these derivatives are singular in the continuum limit. From a Langevin equation perspective where we work with discretized time, the noise at time $t$ is uncorrelated with noise at the next time step, $t+\Delta t$. Therefore the finite difference of noise between different time steps is random, leading to a singular time derivative in the continuum limit.
 While such terms do not affect two‑point functions, they enter at the level of three and higher‑point correlators involving velocity fluctuations. 

This issue can be resolved by working in a hydrodynamic frame in which the energy and momentum fluxes are aligned with the \textit{average} velocity, 
\begin{equation}
    \varepsilon\equiv \bar u_\mu \bar u_\nu T^{\mu\nu},
\quad \pi^\nu \equiv \bar u_\mu \bar \Delta^\nu_{\lambda} T^{\mu\lambda}
\end{equation}
where  
\begin{equation}
    \bar \Delta^{\mu\nu}\equiv g^{\mu\nu}+\bar u^\mu \bar u^\nu.
    \label{eq:bar_delta}
\end{equation} 
In other words, the time-like components of the noise that arise in the Landau frame are absorbed in the definition of energy and momentum flux in the new frame, and and the remaining noise resides purely in the spatial stress sector. Consequently, the evolution equations for equal‑time correlators involve only spatial gradients of noise and are free of singular time derivatives. This frame is introduced in Ref. \cite{An:2026glk} as the \textit{average Landau frame}. 

For Bjorken flow, the average velocity is constant in Bjorken coordinates, $\bar u^\mu=(1,{\bm 0})$ and the associated Lorentz frame with the average Landau frame coincides with the global lab frame. This property follows from the fact that the Bjorken velocity field is conservative, i.e. $\bar u^\mu=\partial_\mu \tau$ and implies the existence of a global space‑like hypersurface orthogonal to the flow. In this special case, the average Landau frame aligns with what is known in the literature as the \textit{density frame} \cite{Armas_2021,de_Boer_2020,Novak_2020}, which has 
been employed recently in stochastic hydrodynamic simulations based on Metropolis algorithm. In this case, the lab frame time is identified with the ``Metropolis time" along which the random Metropolis updates are made \cite{Bhambure:2024axa,Bhambure:2024gnf,Basar:2024qxd}.      
\subsection{Effective action in the density frame}
\label{sec:density_frame}

The effective action for hydrodynamics in the density frame was formulated in \cite{Armas_2021}. It is given as
\begin{equation}
\begin{split}
        \mathcal{L} =& -\varepsilon\nabla_\tau X_a^\tau - (\varepsilon+p)v^i\nabla_i X_a^\tau + w v^i\nabla_\tau X_{ai} + (w v^iv^j+p\delta^{ij})\nabla_j X_{ai} \\
        & - TD^{ijkl}\nabla_k\frac{v_l}{T} \nabla_i X_{aj} + iTD^{ijkl}\nabla_iX_{aj}\nabla_kX_{al}
    \label{eq:density-Lagrangian}
\end{split}
\end{equation}
It can be obtained from the Landau frame expression given in Eq. \eqref{eq:Landau-Lagrangian} via the frame change
\begin{equation}
  w=\gamma^2 w_L,\quad T=T_L/\gamma
  \label{eq:w_T_density}
\end{equation}
where $w=\varepsilon+p$ and $w_L=\varepsilon_L+p$ are enthalpy densities in the density and Landau frames respectively. The fluid four-velocity is $u=\gamma(1,{\bm v})$ where  $ \gamma=1/\sqrt{1-{\bm v}^2}$
is the Lorentz factor. Since pressure is independent of the choice of frame, Eq. \eqref{eq:w_T_density} leads to:
\begin{equation}
    \varepsilon=\gamma^2w_L-p=\gamma^2 \varepsilon_L+\gamma^2 v^2 p
    \label{eq:eps_density}
\end{equation}
Similarly, the entropy in the density frame satisfies
\begin{equation}
    s=\gamma s_L=\frac{w}{T}-\frac{w}{T}v^2\,.
    \label{eq:s_density}
\end{equation}
with
\begin{equation}
    ds=\beta d\varepsilon-\beta {\bm v}\cdot d{\bm \pi}
    \label{eq:ds_density}
\end{equation}
where ${\bm \pi}\equiv w {\bm v}$ is the momentum density. 
Finally the Gibbs-Duhem equation $dp=s_LdT_L$ leads to
\begin{equation}
 \beta dp = -w (1 - v^2) d\beta + \frac{\beta w}{2} dv^2
 \label{eq:density-gibbs-duhem}
\end{equation}
where $\beta=1/T$ as usual. 
The dissipative tensor can be decomposed into components along $v^i$ and orthogonal to $v^i$ as follows. Let $P^{ij}$ be the projection operator orthogonal to $v^i$ and ${\hat v^i}=v^i/|{\bm v}|$ be the unit vector along $v^i$. In general, we can write
\begin{equation}
        D^{ijkl} = D_1(P^{k(i}P^{j)l} - \frac{1}{2}P^{ij}P^{kl})+ D_2 P^{ij}P^{kl} + D_3 \hat{v}^{(i}P^{j)(k}\hat{v}^{l)}+ D_4(P^{ij}\hat{v}^k\hat{v}^l + \hat{v}^i\hat{v}^j P^{kl}) +D_5\,\hat{v}^i\hat{v}^j\hat{v}^k\hat{v}^l 
\end{equation}
where the parentheses in the superscripts denote symmetrization as usual. From the relation between density and Landau frames we deduce that the coefficients, $D_i$s, depend on the shear and bulk viscosities, $\eta$ and $\zeta$, the equation of state, as well as $\gamma$. Their exact forms are listed in section 5.2 of Ref. \cite{Armas_2021}.
As explained above, both dissipative and stochastic terms in the density‑frame action are manifestly spatial. Although the action is not manifestly Lorentz invariant, it is invariant under Lorentz boosts followed by an accompanying redefinition of the hydrodynamic variables.

At this point, it is useful to pause for a moment to contrast the issue with noise we discussed above from the effective action perspective. Unlike the density frame, or more generally the average Landau frame, in the Landau frame, the noise and dissipative terms are spatial in the \textit{fluctuating} fluid frame. In other words they contain projectors  $\Delta^{\mu\nu}$ which has a timelike component in the \textit{average} fluid frame defined by $\bar u$. 
Let us decompose  $\Delta^{\mu\nu}$ as
\begin{equation}
     \Delta^{\mu\nu}= \bar\Delta^{\mu\nu}+\delta u^\mu\bar u^\nu+\bar u^\mu \delta u^\nu+ \delta u^\mu \delta u^\nu,
\end{equation}
where $\bar\Delta^{\mu\nu}$ is defined in Eq.~\eqref{eq:bar_delta}, and focus on the particular term $\bar u^\mu\delta u^\nu$.
This term will generate a time derivative for the noise field at the third-order in the Lagrangian as follows. Let us focus on the shear channel:
\begin{equation}
\begin{split}
    \mathcal{L}_{noise,shear} =iT\eta\Delta^{\mu\nu}\Delta^{\alpha\beta}\nabla_\mu X_{a\alpha}\nabla_\nu X_{a\beta}
   & \supset 
     iT\eta \bar \Delta^{\alpha\beta}(\bar u\cdot \nabla X_{a\alpha})(\delta u\cdot\nabla X_{a\beta})
    \quad \text{(Landau Frame)}
\end{split}
\end{equation}
Therefore, in the stochastic equations there will be a term that involves the time derivative of noise generated by the $\bar u\cdot \nabla X_{a\alpha}$ term. Similar terms exist in the bulk channel as well. In the average Landau frame all such terms are absorbed into the energy-momentum flux leading to the absence of noise in energy and momentum. Noise only appears in the stress channel which is purely spatial.   

\subsection{Dynamical KMS symmetry}
\label{sec:KMS}

Within the EFT formulation, local thermal equilibrium is encoded in the dynamical KMS symmetry, a consequence of the Schwinger–Keldysh construction. This symmetry guarantees the fluctuation–dissipation relation at the level of the effective action \cite{liu-lecture,Glorioso_2017,Jain_2021,Armas_2021}. 
 The dynamical KMS symmetry implies that the density frame Lagrangian given in Eq. \eqref{eq:density-Lagrangian} is invariant under the following transformation of the hydrodynamic fields
\begin{equation}
   \varepsilon(x)\rightarrow \varepsilon(-x) ,\quad v^i(x)\rightarrow v^i(-x),\quad X_{a\tau}(x)\rightarrow -X_{a\tau}(-x)-i\left(\beta(-x)-\beta_0\right)
    ,\quad X_{ai}(x)\rightarrow -X_{ai}(-x)-i\beta_i(-x)
    \label{eq:KMS}
\end{equation}
where   $\beta(x)\equiv 1/{T(x)}$ is the locally varying inverse temperature, $\beta_i(x)\equiv \beta(x)v_i(x)$,
and $\beta_0=1/T_0$ is the inverse temperature in global thermal equilibrium which is constant. It is useful exercise to see how KMS invariance manifests itself. We can write the Lagrangian \eqref{eq:density-Lagrangian}, up to a total derivative, as
\begin{equation}
        \mathcal{L} =X_a^\tau\left(\nabla_\tau\varepsilon  +\nabla_i(wv^i) \right) -X_{ai}\left( \nabla_\tau (w v^i) + \nabla_j (w v^iv^j+p\delta^{ij})\right) + iTD^{ijkl}\nabla_iX_{aj}(\nabla_kX_{al}+i \nabla_k \beta_l ) 
    \label{eq:density-Lagrangian2}
\end{equation}
The variation of the terms up to linear in order $X_a$ reproduces the classical hydrodynamic equations. The last term in Eq. \eqref{eq:density-Lagrangian2}, which includes both dissipative and noise terms, is invariant under the KMS transformation given in Eq. \eqref{eq:KMS}, as a manifestation of the fluctuation-dissipation theorem. The first two terms in Eq. \eqref{eq:density-Lagrangian2} encode ideal hydrodynamics and their variation under KMS transformation leads to, up to a total derivative, 
\begin{equation}
    i\delta_{KMS}{\cal L}=-w(1-v^2) D_t\beta+\frac{1}{2}\beta wD_t( v^2)-\beta D_t p
\end{equation}
where $D_t=\nabla_\tau+{\bm v}\cdot\nabla$ is the convective derivative which vanishes due to the Gibbs-Duhem relation given in Eq. \ref{eq:density-gibbs-duhem}.

\subsection{Effective Lagrangian in the Bjorken background}
\label{sec:bjorken}

To study fluctuations in an expanding medium, we specialize to Bjorken flow, which describes a one‑dimensional, boost‑invariant expansion that is homogeneous and isotropic in the transverse plane.
Let $x_3$ be the expansion axis. Boost invariance means that the hydrodynamic quantities depend only on the proper time, $ \tau\equiv\sqrt{x_0^2-x_3^2}$, and not the space-time rapidity, $y\equiv \text{arctanh}(x_3/x_0)$. It is convenient to use a coordinate system where boost-invariance is manifest such that $g_{\mu\nu} \equiv {\rm diag}(-1,1,1,\tau^2)$ with non vanishing Christoffels  $\Gamma^y_{\tau y }=\Gamma^y_{y\tau}=\frac{1}{\tau}, \Gamma^\tau_{yy}=\tau$. The ideal hydrodynamic equations $\nabla_\mu T^{\mu\nu}=0$ yield  
\begin{equation}
\frac{d\varepsilon}{d\tau}=-\frac{\varepsilon+p}{\tau}=-\frac{w}{\tau}
\label{eq:Bjorken-epsilon-eq}
\end{equation}
For any equation of state, this implies that the entropy density decreases as
\begin{equation}
s(\tau)=s_0\left(\frac{\tau_{\rm in}}{\tau}\right)
\end{equation}
as the plasma expands, which is a consequence of the one-dimensional, boost-invariant expansion. 
Assuming the speed of sound, $c_s$, is constant leads to the well-known expressions 
\begin{equation}
\varepsilon(\tau)=\varepsilon_0\left(\frac{\tau_{\rm in}}{\tau}\right)^{1+c_s^2},\qquad
T(\tau)=T_0\left(\frac{\tau_{\rm in}}{\tau}\right)^{c_s^2}
\label{eq:Bjorken-sol}
\end{equation}
which describe how the plasma cools down as it expands. 

We expand the hydrodynamic fields around the Bjorken background and retain terms up to cubic order in fluctuations. Viscous corrections to the background flow are parametrically suppressed relative to fluctuation effects and are therefore neglected at this order.

In this paper, we focus on the three-point correlators, so it suffices to expand Eq. \eqref{eq:density-Lagrangian} to cubic order in hydrodynamic fields in the Bjorken background.  The fluctuating degrees of freedom are the densities of conserved quantities: energy density  $\delta\epsilon$ and the momentum density $\delta \pi^i\equiv\delta(wu^i)$, which we collectively denote by $\phi^A$. To keep our notation simple, when it is not ambiguous, we will drop the bar when we denote the average quantities and denote $\varepsilon\equiv \bar \varepsilon=\varepsilon_{Bjorken}(\tau)$, $w\equiv\bar w=w_{Bjorken}(\tau)$, and so on for the remainder of the paper. To leading order, the dissipative coefficients are
\begin{equation}
    D_1=2\eta,\quad
    D_2=\zeta+\frac{\eta}{3},\quad
    D_3=4\eta,\quad
    D_4=\zeta-\frac{2\eta}{3},\quad
    D_5=\zeta+\frac{4\eta}{3}
\end{equation}
up to ${\cal O}(\delta v^2)$ corrections which are beyond the order we are working in. As a result, the dissipative tensor takes the familiar form 
\begin{equation}
    D^{ijkl} = \eta(\varepsilon+\delta \varepsilon)(g^{ik}g^{jl} + g^{jk}g^{il}) + \left(\zeta(\varepsilon+\delta \varepsilon) - \frac{2}{3}\eta(\varepsilon+\delta \varepsilon)\right) g^{ij}g^{kl}+{\cal O}(\delta v^2)
\end{equation}
Here $g^{ij}$ denotes the spatial the metric in the lab frame. The viscous coefficients $\eta$ and $\zeta$ above have to be expanded to linear order in $\delta \varepsilon$
\begin{equation}
     \eta(\varepsilon+\delta \varepsilon)=\eta( \varepsilon)+\eta'( \varepsilon)\delta \varepsilon+{\cal O}(\delta \varepsilon^2),\quad 
    \zeta(\varepsilon+\delta \varepsilon)=\zeta(\varepsilon)+\zeta'( \varepsilon)\delta \varepsilon+{\cal O}(\delta \varepsilon^2)
\end{equation}
Consequently, the effective Lagrangian \eqref{eq:density-Lagrangian} up to quadratic order is 
\begin{align}
\begin{split}
\mathcal{L}_2 =&
-\Bigl[\partial_\tau\delta\varepsilon +\frac{1+c_s^2}{\tau}\delta\varepsilon + \partial\cdot\vpi \Bigr]X_{a\tau}
-\Bigl[\partial_\tau\delta\pi^i + \frac{1}{\tau}\delta\pi^i + \frac{1}{\tau}\delta\pi^3\delta^i_{\;3}
+ c_s^2\partial^i\delta\varepsilon\Bigr]X_{ai}
\\
& -\Bigl[\gamma_\eta\bigl(\partial^i\delta\pi^j+\partial^j\delta\pi^i\bigr)
+\bigl(\gamma_\zeta-\tfrac{2}{3}\gamma_\eta\bigr)\,\partial\cdot\vpi\;\delta^{ij}\Bigr]\partial_i X_{aj}
\\
& + iT\Bigl[\eta(\delta^{ik}\delta^{jl}+\delta^{jk}\delta^{il})
+\bigl(\zeta-\tfrac{2}{3}\eta\bigr)\delta^{ij}\delta^{kl}\Bigr]\partial_i X_{aj}\,\partial_k X_{al}
        \label{eq:l2}
    \end{split}
\end{align}
Here, terms in the second and the third line encode the dissipative and noise terms respectively. Note that since the fluid is expanding in the $x_3$ direction, the above expression is only isotropic in the transverse $x_1-x_2$ plane. 
As mentioned earlier, up to second order, there is no difference between density and Landau frames and, as expected, and the effective Lagrangian \eqref{eq:l2} correctly reproduces the equations for the two-point functions derived before in Refs. \cite{akamatsu,An:2019osr}. 
The cubic order Lagrangian is more involved. After some algebra it can be written as
    \begin{align}
    \begin{split}
\mathcal{L}_3 =&
-\Bigl[\frac{p^{\prime\prime}}{2\tau}(\delta\varepsilon)^2-\frac{c_s^2}{\tau w}\,\vpi\cdot\vpi +\frac{1}{\tau w}\delta\pi^3\delta\pi^3    \Bigr]X_{a\tau}
\\
&-\Bigl[\frac{1}{w}\,\vpi\cdot\partial\,\delta\pi^i
+\frac{1}{w}\,\delta\pi^i\,\partial\cdot\vpi
+\frac{1}{2}\,p^{\prime\prime}\,\partial^i(\delta\varepsilon)^2
-\frac{c_s^2}{w}\,\partial^i(\vpi\cdot\vpi)\Bigr]X_{ai}
\\
&+\Bigl[\gamma_\eta\Bigl(\frac{w'}{w}+\frac{T'}{T}\Bigr)\bigl(\partial^i(\delta\varepsilon\,\delta\pi^j)+\partial^j(\delta\varepsilon\,\delta\pi^i)\bigr)
+\Bigl(\gamma_\zeta-\tfrac{2}{3}\gamma_\eta\Bigr)\Bigl(\frac{w'}{w}+\frac{T'}{T}\Bigr)\partial\cdot(\delta\varepsilon\,\vpi)\,\delta^{ij}\Bigr]\partial_i X_{aj}
\\
&-\frac{1}{Tw}\Bigl[(\eta T)'\bigl(\partial^i\delta\pi^j+\partial^j\delta\pi^i\bigr)
+\bigl((\zeta-\tfrac{2}{3}\eta)T\bigr)'\,(\partial\cdot\vpi)\,\delta^{ij}\Bigr]\delta\varepsilon\;\partial_i X_{aj}
\\
&+ i\Bigl[(\eta T)'\,(\delta^{ik}\delta^{jl}+\delta^{jk}\delta^{il})
+\bigl((\zeta-\tfrac{2}{3}\eta)T\bigr)'\delta^{ij}\delta^{kl}\Bigr]\delta\varepsilon\;\partial_i X_{aj}\,\partial_k X_{al}
        \label{eq:l3}
    \end{split}
\end{align}
Notice that in the last line there are two $X_a$ fields and one $\delta\varepsilon$. This is a manifestation of the famous multiplicative noise, where the fluctuations affect the magnitude of the noise. 
  
Having established the effective Lagrangian, it is useful to derive the stochastic equations of motion. We first express the Lagrangian linearly in $X_a$ via a Hubbard-Stratonovich transformation. Its variation with respect to $X_a$ leads to:  
\begin{equation}
\label{eq:deltaeps-evolution}
\partial_\tau \delta\varepsilon =
-\frac{1+c_s^2}{\tau}\,\delta\varepsilon
-\partial\cdot\vpi
-\frac{p^{\prime\prime}}{2\tau}\,(\delta\varepsilon)^2
+\frac{c_s^2}{\tau w}\,\vpi\cdot\vpi - \frac{1}{\tau w}\delta\pi^3\delta\pi^3
\end{equation}
and
\begin{align}
\begin{split}
        \partial_\tau \delta\pi^i
&=
-\frac{1}{\tau}\,\delta\pi^i
-\frac{1}{\tau}\,\delta\pi^3\,\delta^i_{\;3}
-c_s^2\,\partial^i\delta\varepsilon
+\gamma_\eta\,\partial^2\delta\pi^i
+\left(\gamma_\zeta+\frac{\gamma_\eta}{3}\right)\partial^i\partial\cdot\vpi
\\
&\quad
-\frac{1}{w}\,\vpi\cdot\partial\delta\pi^i
-\frac{1}{w}\,\delta\pi^i\partial\cdot\vpi
-\frac{1}{2}\,p^{\prime\prime}\,\partial^i(\delta\varepsilon)^2
+\frac{c_s^2}{w}\,\partial^i\!\bigl(\vpi\cdot\vpi\bigr)
\\
&\quad
-\gamma_\eta\,\left(\frac{w'}{w}+\frac{T'}{T}\right)\,\partial^2(\delta\varepsilon\,\delta\pi^i)
-\left(\gamma_\zeta+\frac{\gamma_\eta}{3}\right)\left(\frac{w'}{w}+\frac{T'}{T}\right)\,
\partial^i\partial\cdot(\delta\varepsilon\,\vpi)
\\
&\quad
+\frac{1}{Tw}\,(\eta T)'\,\partial^l(\delta\varepsilon\,\partial_l\delta\pi^i)
+\frac{1}{Tw}\,(\eta T)'\,\partial_l(\delta\varepsilon\,\partial^i\delta\pi^l)
+\frac{1}{Tw}\,\Bigl(\bigl(\zeta-\tfrac{2}{3}\eta\bigr)T\Bigr)'\,
\partial^i(\delta\varepsilon\partial\cdot\vpi) + \xi^i
\end{split}
\end{align}
with the noise correlator
\begin{equation}
    \ang{\xi^i(\tau_1,\vx_1)\xi^j(\tau_2,\vx_2)} =  -2T\left(\eta\delta^{ij}\partial^2+(\zeta+\frac{1}{3}\eta)\partial^{i_1}\partial^{j_1}\right)\delta^{(3)}(\vx_1-\vx_2)\delta(\tau_1-\tau_2)
\end{equation}
Note that as a result of the density frame construction, there are no dissipative or noise terms in Eq.~\eqref{eq:deltaeps-evolution}. 

Another useful exercise is to see how the KMS invariance manifests in ${\cal L}_2$ and ${\cal L}_3$. Because the Lagrangian \eqref{eq:density-Lagrangian2} already satisfies KMS invariance before expanding, we already know that it must hold for ${\cal L}_2$ and ${\cal L}_3$ as well. But the way it does involves nontrivial cancellations between different terms at  different orders and serves as a cross-check of our calculations.  To make the analysis concise, we only present here the dissipative terms where the spatial gradients contract as $\partial^l f\partial_l g$. Other combination of indices form similar KMS invariant groups. The relevant terms in $\mathcal{L}_2$ and $\mathcal{L}_3$ are
\begin{equation}
\begin{split}
        \mathcal{L}_2+\mathcal{L}_3 \supset 
        &-\gamma_\eta\,\partial^l\delta\pi^i\,\partial_l X_{ai}
        +iT\eta\,\partial^l X_{ai}\,\partial_l X_a^{\,i}
        +\gamma_\eta\Bigl(\frac{w'}{w}+\frac{T'}{T}\Bigr)\,\partial^l(\delta\varepsilon\,\delta\pi^i)\,\partial_l X_{ai}
\\
& -\frac{(\eta T)'}{Tw}\,\delta\varepsilon\,\partial^l\delta\pi^i\,\partial_l X_{ai}
+i(\eta T)'\,\delta\varepsilon\,\partial^l X_{ai}\,\partial_l X_a^{\,i}
\label{eq:Lag-KMS-demo}
\end{split}
\end{equation}
 Here the first two terms come from ${\cal L}_2$ and the remaining terms from ${\cal L}_3$. Under the KMS transformation, expanded to second order in fields, 
\begin{equation}
        X_{ai} \rightarrow -X_{ai} - \frac{i}{Tw}\left[\delta\pi_i - \left(\frac{w'}{w}+\frac{T'}{T} \right)\delta\varepsilon\delta\pi_i  \right],
    \label{eq:KMS-expanded}
\end{equation}
the first two terms in Eq.~\eqref{eq:Lag-KMS-demo}, which come from $\mathcal{L}_2$, combines with the third term and forms a KMS-invariant block. The last two terms are KMS-invariant as a pair. These delicate cancellations between terms that originate from different orders in the Lagrangian provide a useful algebraic cross-check to ensure that the expansion is self-consistent and complete.

\section{Nongaussian Evolution Equations}
\label{sec:evolution}
\subsection{Schwinger-Dyson Equations}
\label{sec:SD}
In this section, we derive deterministic evolution equations for equal‑time two‑ and three‑point correlation functions using the effective Lagrangian constructed in Sec. \ref{sec:generalities}.
 In the EFT formulation, the time evolution of correlation functions follows from the Schwinger–Dyson equations obtained by varying the action with respect to the auxiliary fields,
\begin{equation}
    \big\langle \frac{\delta I}{\delta X_{a\mu}(x_1)}\phi^{A_2} (x_2)\phi^{A_3}(x_3) \big\rangle = 0\,. 
    \label{eq:schwinger-dyson}
\end{equation}
where $I\equiv \int dt {\cal L}$ is the action. 
It is convenient to write the effective Lagrangian in the compact form 
\begin{equation}
{\cal L}=X_{a\mu}E^\mu[\phi]+i X_{a\mu} {\cal Q}^{\mu\nu}[\phi] X_{a\nu} 
\end{equation}
Here $E^\nu=\nabla_\mu T^{\mu\nu}$ encodes the classical equations of motion, and ${\cal Q}^{\mu\nu}$ is the noise tensor, purely spatial in the density frame.
Upon variation with respect to $X_a$, the Schwinger Dyson equations take the form 
\begin{equation}
    \big\langle E^\mu(x_1)\phi^{A_2} (x_2)\phi^{A_3}(x_3) \big\rangle
    +i\big\langle X_{a\nu}(x_1) {\cal Q}^{\mu\nu}(x_1)\phi^{A_2} (x_2)\phi^{A_3}(x_3) \big\rangle = 0 
    \label{eq:DSE}
\end{equation}
Separating time derivatives from spatial fluxes, we write
\begin{equation}
    E^0[\phi]\equiv\partial_\tau (\delta \varepsilon) + f^0[\phi], \quad  E^i[\phi]\equiv\partial_\tau (\delta \pi^i) +f^i[\phi]
\end{equation}
where $f^0[\phi]$ and $f^i[\phi]$, are nonlinear functionals of the hydrodynamic fields and their spatial gradients. Their explicit forms follow from the quadratic and cubic Lagrangians given in Eqs. \eqref{eq:l2} and \eqref{eq:l3}. The Schwinger-Dyson equations in the density frame then become
\begin{eqnarray}
    \partial_{\tau_1} \langle \delta e(x_1) \phi^B(x_2)\phi^C(x_3)\rangle &=& -\langle f^{0}(x_1) \phi^{A_2} (x_2)\phi^{A_3}(x_3)\rangle
    \label{eq:eps-equation}
    \\
      \partial_{\tau_1} \langle \delta \pi^i(x_1) \phi^{A_2} (x_2)\phi^{A_3}(x_3)\rangle &=& -\langle f^i(x_1) \phi^{A_2} (x_2)\phi^{A_3}(x_3)\rangle-i\left\langle X_{a j}(x_1) {\cal Q}^{ij}(x_1)\phi^{A_2} (x_2)\phi^{A_3}(x_3) \right\rangle
      \label{eq:pi-equation}
\end{eqnarray}
To leading order in the hydrodynamic limit,
$f^0$ and $f^i$ need only be expanded to quadratic order in fluctuations. Linear terms generate contributions proportional to three‑point functions, while quadratic terms produce products of two‑point functions via Wick contractions.

The noise kernel ${\cal Q}^{ij}$ must be expanded only to linear order in fluctuations, since it enters the action quadratically in $X_a$. Expanding ${\cal Q}^{ij}$ to zeroth order, we get $\langle X_a\phi\phi\rangle$ terms, for which we can use the general Schwinger-Keldysh relation  \cite{crossley2017effectivefieldtheorydissipative}
\begin{equation}
    \begin{split}
        \langle X_{a}^i(\tau_1)\phi^{B}(\tau_2)\phi^{C}(\tau_3) \rangle  &= 
        \frac{1}{2}\Theta(\tau_{31})\Theta(\tau_{21})\langle[
        \{\phi^{B},\phi^{C}\},X_a^i] 
        \rangle 
        + \frac{1}{2}\Theta(\tau_{31})\Theta(\tau_{12})\langle 
        \{[\phi^C,X_a^i],\phi^B \}\rangle 
        \\&+ \frac{1}{2}\Theta(\tau_{13})\Theta(\tau_{21})\langle 
        \{[\phi^B,X_a^i],\phi^C\} \rangle\,.
    \end{split}
\end{equation}
where $\Theta$ denotes the Heaviside step function and $\tau_{ij}\equiv \tau_i-\tau_j$. Here we only explicitly denoted the temporal dependence to keep our notation simple. The $X_a$ dependence can be eliminated by noticing that $X_{aA}$ is the conjugate momentum of $\phi^A$, therefore
\begin{equation}
    [X_a^i(\vx_1,\tau),\delta \pi^j(\vx_2,\tau)] = i\delta^{ij}\,\delta^{(3)}(\vx_1-\vx_2)
\end{equation}
As a result we see that all $\langle X_a\phi\phi\rangle $ correlators vanish. Expanding ${\cal Q}^{ij}$ to first order leads to $\langle X_a\phi\phi\phi\rangle$ terms. To leading order in the fluctuation expansion only the disconnected part of them, proportional to $ G_2\times G_2\sim \epsilon^2$, contribute. The non-vanishing Wick contractions are  
\begin{eqnarray}
\langle X_{ai}(\vec{x}_1)\,\delta\varepsilon(\vec{x}_1)\,\delta\pi^{j}(\vec{x}_2)\,\phi^{A}(\vec{x}_3) \rangle
&=&
\langle X_{ai}(\vec{x}_1)\,\delta\pi^{j}(\vec{x}_2) \rangle \,
\langle \delta\varepsilon(\vec{x}_1)\,\phi^{A}(\vec{x}_3)\rangle
+
\langle X_{ai}(\vec{x}_1)\,\phi^{A}(\vec{x}_3) \rangle \,
\langle \delta\varepsilon(\vec{x}_1)\,\delta\pi^{j}(\vec{x}_2)\rangle
\\
&=&
-\frac{i}{2}\,\delta^{j}_{i}\,\delta^{(3)}(\vec{x}_1-\vec{x}_2)\,G^{\varepsilon A}(x_1,x_3)
-\frac{i}{2}\,\delta^{A}_{i}\,\delta^{(3)}(\vec{x}_1-\vec{x}_3)\,G^{\varepsilon j}(\vec{x}_1,\vec{x}_2)
\end{eqnarray}
where all the terms above are evaluated at equal-time $\tau$.

Putting everything together, we arrive at the evolution equations for the connected, equal-time two and three-point functions: 
\begin{align}
\begin{split}
\label{eq:G2-evolution}
        &\partial_\tau G^{A_1A_2}(\vx_1,\vx_2;\tau) = 2\left[L^{A_1}_{B}(\vx_1;\tau)G^{BA_2}(\vx_1,\vx_2;\tau) + Q^{A_1A_2}(\vx_1,\vx_2;\tau)\right]_{\overline{12}} 
\end{split} \\
\begin{split}
\label{eq:G3-evolution}
        &\partial_\tau G^{A_1A_2A_3}(\vx_1,\vx_2,\vx_3;\tau) = 3\left[ L^{A_1}_{B}(\vx_1;\tau)G^{BA_2A_3}(\vx_1,\vx_2,\vx_3;\tau) + 2L^{A_1}_{BC}(\vx_1;\tau) G^{B A_2}_c(\vx_1,\vx_2;\tau) G^{C A_3}(\vx_1,\vx_3;\tau) \right.\\
        & \qquad\qquad\qquad\qquad + \left. 2Q^{A_1A_3}_{\varepsilon}(\vx_1,\vx_3;\tau)G^{\varepsilon A_2}(\vx_1,\vx_2;\tau) \right]_{\overline{123}}
\end{split}
\end{align}
where the multi-linear operators $L$ and $Q$ are given in Appendix \ref{sec:operators}. Following Ref. \cite{An:2022jgc} we used the notation $[\dots]_{\overline{1\dots n}}\equiv1/n![\dots]_{{\rm Perm}(1\dots n)}$ for normalized permutations. Note that both the indices of the correlators and their associated coordinates permute under the permutation symbol. 

\subsection{Multi-point Wigner function}

To analyze the evolution equations in wave-vector space, it is useful to perform a generalized Wigner transform of the fluctuation correlators. Following Ref. \cite{An:2020vri}, we define the Fourier transform with respect to relative coordinates while retaining dependence on the slow “center‑of‑mass” coordinate:
\begin{align}
     &W^{A_1\dots A_N}(\vq_1,...,\vq_n;\vx,\tau) \equiv \int d^3\vy_1...d^3\vy_N\,G^{A_1\dots A_N}(\vx+\vy_1,...,\vx+\vy_N;\tau)\,\delta^{(3)}\left(\frac{\vy_1+...\vy_N}{N}\right)e^{-i(\vq_1\cdot\vy_1+...+\vq_N\cdot\vy_N)} 
     \label{wigner}
     \\
     &G^{A_1\dots A_N}(\vx_1,...,\vx_N;\tau) = \int \frac{d^3\vq_1}{(2\pi)^3}...\frac{d^3\vq_N}{(2\pi)^3}W^{A_1\dots A_N}(\vq_1,...,\vq_N;\vx,\tau)(2\pi)^3\delta^{(3)}(\vq_1+...+\vq_N)e^{i(\vq_1\cdot\vx_1+...+\vq_N\cdot\vx_N)}
     \label{inverse-wigner}
\end{align}  
The Wigner transform is invariant under a constant shift for  of all the $\vq_i$s, i.e $\vq_i\rightarrow \vq_i+const$. This means there are only $N-1$ independent fluctuation wave-vectors. The remaining degree of freedom, associated with the ``center of fluctuation" coordinate $\vx=\sum_i\vx_i/N$ and captures the slower background motion, is not Fourier transformed. To keep the expressions simple we set the constant as $\sum_i \vq_i=0$. With the above conventions, the Wigner transform of the correlation functions appear in the equations for the two and three-point equations for Bjorken flow are:
\begin{equation}
    \begin{split}
        &G_2(\vx_1,\vx_2) \rightarrow{} W_2(\vq,-\vq) \\
        &G_3(\vx_1,\vx_2,\vx_3) \rightarrow{} W_3(\vq_1,\vq_2,-\vq_1-\vq_2) \\
        &G_2(\vx_1,\vx_2)G_2(\vx_1,\vx_3) \rightarrow{} W_2(-\vq_2,\vq_2)W_2(-\vq_3,\vq_3)
    \end{split}
\end{equation}
Note that for arbitrary background flow which does not necessarily carry the symmetries of Bjorken flow, these relations have to be replaced with substantially more complicated expressions \cite{An:2026glk}.

In the wave-vector space, the evolution equations take the form 
\begin{align}
\begin{split}
   & \partial_\tau W^{A_1A_2}(\vq_1,\vq_2;\tau) = 2 \left[ L^{A_1}_{B}(
    \vq_1)W^{B A_2}(\vq_1,\vq_2;\tau) + Q^{A_1A_2}(\vq_1,\vq_2) \right]_{\overline{12}}
    \label{eq:2pt}
\end{split} \\
\begin{split}
    &\partial_\tau W^{A_1A_2A_3}(\vq_1,\vq_2,\vq_3;\tau)= 3\left[ L^{A_1}_{B}(\vq_1)W^{B A_2 A_3}(\vq_1,\vq_2,\vq_3;\tau)+2Q^{A_1A_2}_{\varepsilon}(\vq_1,\vq_2,-\vq_3)W^{\varepsilon A_3}(-\vq_3,\vq_3;\tau) \right.\\
    &\phantom{\partial_\tau W^{A_1A_2A_3}(\vq_1,\vq_2,\vq_3;\tau)=}\quad \,\,\,
    \left. + 2L^{A_1}_{BC}(\vq_1,-\vq_2,-\vq_3) W^{B A_2}(-\vq_2,\vq_2;\tau) W^{C A_3}(-\vq_3,\vq_3;\tau)\right]_{\overline{123}}
    \label{eq:3pt}
\end{split}
\end{align}
Similar to the coordinate space, the repeated indices are summed over, and both the indices and their associated wave-vectors permute under the permutation symbols. The non-vanishing coefficients, the Wigner transforms of the multi-linear operators given in Appendix \ref{sec:operators}, are given as 
\begin{align}
    \begin{split}
    \label{eq:L2}
&L^{\varepsilon}_{\varepsilon}(\vq_1)=  -\frac{1+c_s^2}{\tau}
\;\;,\;\;
L^{\varepsilon}_{i}(\vq_1)= -i q_{1i}
\;\;,\;\;
L^{i}_{\varepsilon}(\vq_1)= -i c_s^2 q_1^{\,i}
\\
&L^{i}_{j}(\vq_1)= -\delta^i_{\;j}\frac{1}{\tau}
-\delta^i_{\;3}\delta^3_{\;j}\frac{1}{\tau}
-\delta^i_{\;j}\gamma_\eta\,\vq_1^{\,2}
-\left(\gamma_\zeta+\frac{\gamma_\eta}{3}\right) q_1^{\,i}q_{1j}
\end{split}
\\
\begin{split}
   \label{eq:L3}
&L^{\varepsilon}_{\varepsilon\varepsilon}(\vq_1,-\vq_2,-\vq_3)=  -\frac{p^{\prime\prime}}{2\tau}
\;\;,\;\;
L^{\varepsilon}_{ij}(\vq_1,-\vq_2,-\vq_3)= \delta_{ij}\,\frac{c_s^2}{\tau w} -\delta^3_i\delta^3_j\frac{1}{\tau w}
\;\;,\;\;
L^{i}_{\varepsilon\varepsilon}(\vq_1,-\vq_2,-\vq_3)= -\frac{i}{2}\,p^{\prime\prime}\,q_1^{\,i}
\\
&L^{i}_{jk}(\vq_1,-\vq_2,-\vq_3)
 = -\frac{i}{w}q_{1j}\delta^{i}_k+\frac{i}{w}c_s^2q_1^i\delta_{jk}
\\
&L^{i}_{\varepsilon j}(\vq_1,-\vq_2,-\vq_3)
=  \gamma_\eta\,\left(\frac{w'}{w}+\frac{T'}{T}\right)\,\delta^i_{\;j}\,\vq_1^{\,2}
+\left(\gamma_\zeta+\frac{\gamma_\eta}{3}\right)\left(\frac{w'}{w}+\frac{T'}{T}\right)\,q_1^{\,i}q_{1j}
\\
&
+\frac{(\eta T)'}{Tw}\Bigl[
\delta^i_{\;j}\,(\vq_1\cdot\vq_3)
+ q_{1j}q_3^{\,i}
+ q_1^{\,i}q_{3j}
\Bigr]
+\frac{\Bigl(\bigl(\zeta-\tfrac{2}{3}\eta\bigr)T\Bigr)'}{Tw}\,q_1^{\,i}q_{3j}
\end{split}
 \\
\begin{split}
   \label{eq:Q}
&Q^{ij}(\vq_1,\vq_2) =  \delta^{ij}\frac{Tw}{\tau}\gamma_\eta\vq_1^2 + \frac{Tw}{\tau}\left(\gamma_\zeta+\frac{1}{3}\gamma_\eta\right) q_{1i}q_{1j} 
\\
 &Q^{ij}_{\varepsilon}(\vq_1,\vq_2,-\vq_3) =  - \delta^{ij}\frac{(T\eta)'}{\tau}\vq_1\cdot\vq_2 - \frac{(T\eta)'}{\tau}q_1^jq_2^i - \frac{(T(\zeta-\frac{2}{3}\eta))'}{\tau}q_1^iq_2^j
    \end{split}
\end{align}
The advantage if working in Wigner space is that these coefficients are not differential operators. As a cross-check of our method, we verified that the two-point equation given in Eq. \eqref{eq:2pt} agrees with the one calculated in Ref. \cite{akamatsu} as expected.

These equations can be solved hierarchically. The background Bjorken flow serves as the one‑point function, which is an input for the two‑point function equations. The solutions of the two-point functions then act as sources for the evolution of three‑point functions. In Section \ref{sec:analytical} we present the analytical solutions of these equations, their late-time behavior, as well as numerical illustrations for a select set of parameters.

\subsection{Stationary solutions and equilibrium}

Before solving the evolution equations dynamically, it is instructive to verify that they admit the correct equilibrium solutions. Since the effective action satisfies dynamical KMS symmetry, the evolution equations must reproduce equilibrium correlation functions as stationary solutions.
 
In the density frame, equilibrium correlators follow from the entropy functional,
\begin{equation}
    S[\phi] = \int d^3x(s - \bar{\beta}\varepsilon),
\end{equation}
where $s(\varepsilon,{\bm \pi})$ is the entropy density in the density frame and $\bar{\beta}$ is a Lagrange multiplier that ensures energy conservation.  
The equilibrium fluctuations can be calculated via the path integral where $-S[\phi]$ plays the role of the Euclidean action:   
\begin{equation}
    \langle \dots \rangle_{\rm eq} =\frac{1}{Z}\int {\cal D}\phi\, e^{S[\phi]} (\dots) 
\end{equation}
Expanding $S[\phi]$ to cubic order in fields we obtain
\begin{equation}
    S=\int d^3x\left[s_0+\left(\frac{\partial s}{\partial\varepsilon}-\bar\beta\right)\delta \varepsilon-\frac{\beta  c_s^2}{2 w}\delta \varepsilon ^2 -\frac{ \beta}{2 w} \delta\bm\pi\cdot  \delta \bm \pi +\frac{1}{6}\frac{\partial ^3s}{\partial \varepsilon ^3}\delta \varepsilon ^3+\frac{\beta  (2 c_s^2+1)  }{2 w^2}\delta\varepsilon \delta\bm\pi\cdot  \delta \bm \pi +{\cal O}(\phi^4) \right].
\end{equation}
The equilibrium values of the two and three functions then follow straightforwardly from evaluating the Gaussian integrals
\begin{equation}
\label{eq:Weq}
   W^{\varepsilon\varepsilon}_{\rm eq} = \frac{Tw}{c_s^2\tau}, \quad W^{ij}_{\rm eq} = \frac{Tw}{\tau}\delta^{ij},\quad 
  W^{\varepsilon\varepsilon\varepsilon}_{\rm eq}=\frac{T^2 w}{\tau^2}\left(\varepsilon_p^2+2\varepsilon_p+w\varepsilon_{pp}\right),\quad 
  W^{\varepsilon ij}_{\rm eq}=W^{i\varepsilon j}_{\rm eq}=W^{ij\varepsilon}_{\rm eq} = (1+2c_s^2)\frac{T^2w}{c_s^2\tau^2}\delta^{ij}
\end{equation}
The factors of $1/\tau$ are due to the normalization of the spatial delta functions which follow from $\sqrt{-g}=\tau$. The equilibrium correlators are local in position space and, equivalently, their Wigner transforms given in Eq.~\eqref{eq:Weq} are independent of the wave-vectors ${\bm q}_i$. 
The rest of the two and three-point correlation functions not listed in Eq.~\eqref{eq:Weq} vanish in equilibrium. 

As an illustration we show how $W_{\rm eq}^{ij\varepsilon}$, and $W_{\rm eq}^{ijk}$ satisfy the evolution equations. Similarly to the KMS check, this will serve as an independent cross-check of the evolution equations. 

For the $W_{\rm eq}^{ij\varepsilon}$ correlator we only present the calculation involving the following terms
\begin{equation}
    \begin{split}
        \partial_\tau W_{\rm eq}^{ij\varepsilon}(\vq_1,\vq_2,\vq_3;\tau) 
        \supset & -(\vq_1^2+\vq_2^2)\gamma_\eta W^{ij\varepsilon}_{\rm eq} + \left(\vq_1^2+\vq_2^2\right)\gamma_\eta\left(\frac{w'}{w}+\frac{T'}{T}\right)W^{ij}_{\rm eq}W^{\varepsilon\varepsilon}_{\rm eq} \\
        & + 2\frac{(\eta T)'}{Tw} \vq_1\cdot\vq_2\delta^{ij}W^{li}_{\rm eq}W^{\varepsilon\varepsilon}_{\rm eq} -2(\eta T)'\vq_1\cdot\vq_2\delta^{ij}W^{\varepsilon\varepsilon}_{\rm eq}
    \end{split}
\end{equation}
The remaining terms with different index structure for the wave-vectors cancel in a similar way. Substituting the equilibrium values of the correlation functions, we see that the terms in the first line cancel each other \footnote{Here the following thermodynamic identities are useful: $w'=c_s^2+1$ and $T'=c_s^2 T/w$}
Similarly, the first term in the second line that involves the product of the two-point functions cancels against the last term that has a single two-point function. The former term comes from expanding the dissipative term to quadratic order in fields. More precisely, the $f^i\phi\phi$ term in Eq.~\eqref{eq:pi-equation}. Whereas the latter terms comes from expanding the noise matrix, ${\cal Q}$, in the $X_A{\cal Q}\phi\phi$ term in Eq.\eqref{eq:pi-equation} to linear order in $\phi$. This expansion is a manifestation of multiplicative noise.  

The momentum three-point function $W^{ijk}$ is also interesting. Due to isotropy it must vanish in equilibrium: $W^{ijk}_{\rm eq}=0$. For the same reason we also have $W^{\varepsilon i}_{\rm eq}=0$. Therefore in equilibrium, Eq.~\eqref{eq:3pt} for $W^{ijk}$ reduces to
\begin{equation}
    \partial_\tau W^{ijk}_{\rm eq}(\vq_1,\vq_2,\vq_3;\tau)
    = 3\Big[ L^i_{\;\varepsilon}(\vq_1)W^{\varepsilon jk}_{\rm eq}(\vq_1,\vq_2,\vq_3;\tau)
    + 2L^i_{\;lm}(\vq_1,-\vq_2,-\vq_3)W^{lj}_{\rm eq}(-\vq_2,\vq_2;\tau)W^{mk}_{\rm eq}(-\vq_3,\vq_3;\tau) 
    \Big]_{\overline{123}}\,
    \label{eq:w3-eql}
\end{equation}
which must vanish. Notice that in this equation there are no dissipative/noise terms. Therefore different terms that involve two and three-point functions have to cancel purely on kinematic grounds. In equilibrium, the correlators are $\vq$ independent. However the $L$ coefficients do depend on $\vq_i$s. Using Eqs.~\eqref{eq:L2}, \eqref{eq:L3}, and \eqref{eq:Weq} we find  
\begin{eqnarray}
    \partial_\tau W^{ijk}_{\rm eq}(\vq_1,\vq_2,\vq_3) &= &W_{3\,\rm eq} (-i c_s^2) \left[ q^i_1\delta^{jk}
     +q^j_2\delta^{ik}
   +  q^k_3\delta^{ij}
    \right]  +  
   \left(W_{2\,\rm eq}\right)^2  \frac{i}{w}\left[ 
    -q^j_1\delta^{ik}+c_s^2q_1^i\delta^{jk}
  -q^k_1\delta^{ij}+c_s^2q_1^i\delta^{jk}
    \right.
   \nonumber \\&&
    \left.
     -q^i_3\delta^{jk}+c_s^2q_3^k\delta^{ij}
   -q^i_2\delta^{jk}+c_s^2q_2^j\delta^{ik}
   -q^k_2\delta^{ij}+c_s^2q_2^j\delta^{ik}
   -q^j_3\delta^{ik}+c_s^2q_3^k\delta^{ij}
      \right] 
\end{eqnarray}
where $W_{3\,\rm eq}\equiv (1+2c_s^2)T^2w/(c_s^2\tau^2)$ and $W_{2\,\rm eq}\equiv Tw/\tau$. We also explicitly wrote down all the permutations for clarity. Collecting the terms with the same indices, we see that 
\begin{eqnarray}
    \partial_\tau W^{ijk}_{\rm eq}(\vq_1,\vq_2,\vq_3) &= & (1+2c_s^2)\frac{T^2w}{c_s^2\tau^2}(-ic_s^2) \left[ q^i_1\delta^{jk}
     +q^j_2\delta^{ik}
   +  q^k_3\delta^{ij}
    \right]
   \nonumber \\&&
    +  \frac{i}{w}
   \left(\frac{Tw}{\tau}\right)^2 \left[ 
   - (q^j_1+q^j_3)\delta^{ik}
  -(q^k_1+q^k_2)\delta^{ij}  -(q^i_3+q^i_2)\delta^{jk}+2c_s^2(q_1^i\delta^{jk}+q_1^i\delta^{jk}+q_3^k\delta^{ij})
      \right] \qquad
      \\
&=&-i\frac{T^2w}{\tau^2} \left[(q^i_1+q^i_2+q^i_3)\delta^{jk}+(q^j_1+q^j_2+q^j_3)\delta^{ik}+(q^k_1+q^k_2+q^k_3)\delta^{ij}\right]=0
\end{eqnarray}
since $\vq_1+\vq_2+\vq_3=0$. These highly nontrivial cancellations among different terms involving equilibrium values of $W_3$ and $W_2^2$, as well as kinematical factors provide another nontrivial cross-check for the evolution equations.

\subsection{Averaging Out the Fast Modes}

In the hydrodynamic limit, the typical fluctuation time scale is much shorter than the expansion time scale, $c_s |\vq| \gg 1/\tau$, leading to a separation between fast oscillatory modes with $\omega \sim c_s |\vq|$ and slow relaxational modes with $\omega\sim-i\geta |\vq|^2$ which is of the order the expansion scale $1/\tau$. Diagonalizing
\begin{equation}
L^A_{\;B} = 
\begin{pmatrix}
    0 & \vq  \\
    c_s^2\vq & 
\end{pmatrix}+{\cal O}(|{\vq}|^2,1/\tau),
\end{equation}
yields the eigenvalues 
\begin{equation}
    \lambda_\pm = \pm c_s|\vq|, \quad \lambda_{\perp}=0+{\cal O}(|{\vq}|^2,1/\tau), 
\end{equation}
where the fast modes with $\omega\sim\lambda_{\pm}$ exhibit fast oscillations and throughout the expansion of the fluid would average out to zero. The relaxation modes $\omega\sim \lambda_{\perp}\sim -i \geta|\vq|^2$ dominate the long-time behavior. The eigenvectors of $L$ form a 4 x 4 matrix
\begin{equation}
\label{eq:U}
    U^A_{\;\;\alpha} \equiv 
\begin{pmatrix}
    \frac{1}{\sqrt{2}} & \frac{1}{\sqrt{2}} & 0 & 0     \\
    -\frac{c_s}{\sqrt{2}}\hat{\vq} & \frac{c_s}{\sqrt{2}}\hat{\vq} & \vt_1 &\vt_2\,.
\end{pmatrix}
\end{equation}
Here $\vt_1$ and $\vt_2$ are orthonormal basis vectors perpendicular to $\hat \vq$. In general, the multi-point functions such as three-point functions are characterized by multiple wave-vectors. 
The slow modes can be singled out by expressing the modes in a basis where $L$ is diagonal:
\begin{eqnarray}
     W^{\alpha\beta}(\vq_1,\vq_2;\tau) &=& U^\alpha_{\;\;A}(\vq_1)U^\beta_{\;\;B}(\vq_2)W^{AB}(\vq_1,\vq_2;\tau)
\end{eqnarray}
where $\alpha$ and $\beta$ takes values $\{+,-,[1],[2]\}$, representing directions defined by eigenvectors given in Eq.\eqref{eq:U}. The square brackets indicate orthogonality to $\vq$. The slow modes are $W^{+-}$,$W^{-+}$ and $W^{[ab]}$ where $a,b={1,2}$. Refs. \cite{akamatsu,martinez_2019,An:2019osr,An:2019csj} showed that these modes are related to the distribution functions of phonon (upon-rescaling) whose evolution equations,
\begin{eqnarray}
\label{eq:W2-pm}
        \partial_\tau W^{\pm\mp}(\vq;\tau) &=& -\frac{\zeta+\frac{4}{3}\eta}{w}\vq^2\left[W^{\pm\mp}(\vq;\tau) - \frac{Tw}{\tau}\right] - \frac{1}{\tau}(2+c_s^2 + \cos^2\theta)W^{\pm\mp}(\vq;\tau)    \\
        \label{eq:W2-11}
        \partial_\tau \w{1}{1}(\vq;\tau)  & =& -2\geta\vq^2\left[\w{1}{1}(\vq;\tau)  -\frac{Tw}{\tau}\right] - \frac{2}{\tau}\w{1}{1}(\vq;\tau)    \\
         \label{eq:W2-22}
        \partial_\tau \w{2}{2}(\vq;\tau)  & =& -2\geta\vq^2\left[\w{2}{2}(\vq;\tau)   - \frac{Tw}{\tau}\right] - \frac{2}{\tau}(1+\sin^2\theta)\w{2}{2}(\vq;\tau)   \\
         \label{eq:W2-12}
        \partial_\tau \w{1}{2}(\vq;\tau)   &=& -2\geta\vq^2 \w{1}{2}(\vq;\tau)   - \frac{2}{\tau}(1+\frac{1}{2}\sin^2\theta)\w{1}{2}\,,(\vq;\tau)  
    \end{eqnarray}
correspond to kinetic equations that evolve in the background of the fluid. Note that the longitudinal $+-$ and $-+$ modes have dispersion relations $\omega_{+-}= c_s(+|\vq_1|-|\vq_2|)$ and $\omega_{-+}= c_s(-|\vq_1|+|\vq_2|)$ respectively. Because for the two-point functions we have the constraint $\vq_1+\vq_2=0$, $\omega_{\pm \mp}=0$ and these longitudinal modes are slow. 

Similarly, we can isolate the slow modes for the three-point functions via 
    \begin{equation}
        W^{\alpha\beta\gamma}(\vq_1,\vq_2,\vq_3;\tau) = U^\alpha_{\;A}(\vq_1)U^\beta_{\;B}(\vq_2)U^\gamma_{\;\;C}(\vq_3)W^{ABC}(\vq_1,\vq_2,\vq_3;\tau)\,.
    \end{equation}
In contrast with the two-point functions, the only slow modes here are purely transverse, $W^{[abc]}(\vq_1,\vq_2,\vq_3)$, with $a,b,c=1,2$. It is also important to keep in mind that the $a,b,c$ indices are correlated with the wave-vector indices. Namely, $W^{[abc]}(\vq_1,\vq_2,\vq_3;\tau)=t^{[a]}_i(\vq_1)t^{[b]}_j(\vq_2)t^{[c]}_k(\vq_3) W^{i,j,k}(\vq_1,\vq_2,\vq_3\tau)$. This is the reason why there are no slow longitudinal modes for three-point functions. All the longitudinal modes oscillate with linear combinations of $|\vq_1|$, $|\vq_2|$ and $|\vq_3|$ which do not vanish in general except for specific corner cases in phase space which are measure zero. Therefore the only slow modes are transverse.    

The transverse modes satisfy the following evolution equation:
\begin{equation}
\begin{split}
        \partial_\tau\W{a}{b}{c}&(\vq_1,\vq_2,\vq_3;\tau) = 3\big[ \Lang{a}{d}(\vq_1)\W{d}{b}{c}(\vq_1,\vq_2,\vq_3;\tau) \\
        & + 2\LLang{a}{d}{e}(\vq_1,-\vq_2,-\vq_3)\w{d}{b}(-\vq_2,\vq_2;\tau)\w{e}{c}(-\vq_3,\vq_3;\tau)  \big]_{\overline{123}}
\end{split}
\label{eq:W3-eqn}
\end{equation}
with the operators being
\begin{eqnarray}
\Lang{a}{b}(\vq) &=& 
-\left[ \geta\vq^2+\frac{1}{\tau}\left(1+ t^{[a]}_3 t^3_{[b]}\right) \right]\delta^{[a]}_{\;[b]}
\\
   \LLang{a}{b}{c}(\vq_1,-\vq_2,-\vq_3) &=& \frac{a}{w}\big[\vt^{[a]}(\vq_1)\cdot\vt_{[c]}(\vq_3)\big] \big[\vt_{[b]}(\vq_2)\cdot\vq_{3}\big]
    \label{eq:L3-slow}
\end{eqnarray}

In the next section we present the analytical solutions to these equations along with numerical demonstrations.

\section{Analytical solutions for Bjorken Flow}
\label{sec:analytical}

\subsection{Two-point correlators}

Owing to the simplicity of the Bjorken background, the fluctuation equations derived in Sec.~\ref{sec:evolution} reduce to first‑order ordinary differential equations that admit analytical solutions. In this section, we present these solutions and illustrate their properties for representative choices of parameters and equations of state.

In the hydrodynamic limit, the characteristic expansion time scale, controlled by $c_s\tau$, is much larger than the intrinsic fluctuation time scale $1/q$.  Consequently, it is sufficient to keep only the ideal contribution to the background Bjorken flow, while viscous effects enter through the fluctuation dynamics since $\gamma_\eta|\vq|^2\sim1/\tau$. In this limit, the equilibrium value of the two‑point function takes the simple form
\begin{equation}
    W_{2\, \rm eq}(\tau)=\frac{T(\tau) w(\tau)}{\tau}= W_{2\, \rm eq}(\tau_{\rm in}) \left(\frac{\tau_{\rm in}}{\tau}\right)^{2+2c_s^2}\,.
\end{equation}
For notational convenience, we introduce the dimensionless proper time $\hat\tau\equiv\tau/\tau_{\rm in}$. Assuming isotropic initial conditions, such that  $\w{1}{2}(\bm q,\tau)=0$, the evolution equations, \eqref{eq:W2-pm}-\eqref{eq:W2-12},  can be integrated analytically:
\begin{align} 
     \begin{split}
     \label{eq:W11-soln}
\w{1}{1}(\vq;\tau)=&\frac{\w{1}{1}(\tau_{\rm in})}{\hat\tau^2} \,
e^{-2 \left(\Gamma(\vq^2,\hat\tau)  -\Gamma(\vq^2,1) \right)} 
\\
& +  \frac{2 \Gamma(\vq^2,1)}{\hat\tau^2} W_2^{\rm eq}(\tau_{\rm in})  \,e^{-2 \Gamma(\vq^2,\hat\tau)} 
     \left[
    E_{\kappa_1}\left(-2 \Gamma(\vq^2,1) \right) 
      -\hat\tau^{1-c_s^2}
       E_{\kappa_1}\left(-2 \Gamma(\vq^2,\hat\tau)\right)
       \right]\qquad 
\end{split}
\\
\begin{split}
 \label{eq:W22-soln}
\w{2}{2}(\vq;\tau)=&\frac{\w{2}{2}(\tau_{\rm in}) }{\hat\tau^{2+2\sin^2\theta}} 
\,e^{-2 \left(\Gamma(\vq^2,\hat\tau)-\Gamma(\vq^2,1)\right)} 
 \\
&+ 
   \frac{2 \Gamma(\vq^2,1)}{\hat\tau^{2+2\sin^2\theta}} W_2^{\rm eq}(\tau_{\rm in})\,  e^{-2 \Gamma(\vq^2,\hat\tau)} 
     \left[
    E_{\kappa_2}\left(-2 \Gamma(\vq^2,1) \right) 
      -\hat\tau^{1-c_s^2+2\sin^2\theta}
       E_{\kappa_2}\left(-2 \Gamma(\vq^2,\hat\tau)\right)
       \right]\qquad 
\end{split}       
       \\
\begin{split} 
 \label{eq:Wpm-soln}
W^{\pm\mp}(\vq;\tau)=&\frac{W^{\pm\mp}(\tau_{\rm in})}{\hat\tau^{2+c_s^2+\cos^2\theta}} 
e^{-\frac{4}{3} \left(\Gamma(\vq^2,\hat\tau)  -\Gamma(\vq^2,1) \right)}  
\\
&+ 
   \frac{4 \Gamma(\vq^2,1)}{3\hat\tau^{2+c_s^2+\cos^2\theta}} W_2^{\rm eq}(\tau_{\rm in})  e^{-\frac{4}{3} \Gamma(\vq^2,\hat\tau)} 
     \left[
    E_{\kappa_\pm}\left(-\frac{4}{3} \Gamma(\vq^2,1) \right) 
      -\hat\tau^{1+\cos^2\theta}
       E_{\kappa_\pm}\left(-\frac{4}{3} \Gamma(\vq^2,\hat\tau)\right)
       \right]\qquad 
 \end{split}
\end{align}
where
\begin{equation}
    \kappa_1\equiv \frac{2 c_s^2}{\csp},\quad
    \kappa_2\equiv \frac{2 c_s^2-2\sin^2\theta}{\csp},\quad
    \kappa_\pm\equiv \frac{2 c_s^2-\cos^2\theta}{\csp},
    \quad w'=1+c_s^2
\end{equation}
and $E_n(x)=\int_1^\infty e^{-tx}t^{-n} dt$ is the usual exponential integral. The $\vq$ dependent relaxation rate is given as
\begin{equation}
\Gamma(\vq^2,\hat\tau) \equiv \int_0^{\hat\tau} {\vq^2} \gamma_\eta ( \tau_{\rm in} u)du =   \frac{1}{\csp} \frac{(\eta/s)\tau_{\rm in}}{T(\tau_{\rm in})} \vq^2\hat\tau^{\csp}\,.
\end{equation}
this expression makes it clear that the longer wavelength fluctuations approach equilibrium slower. 
Note that we assumed, for simplicity, a constant value of $\eta/s$.

   The solutions given in Eqs.\eqref{eq:W11-soln}, \eqref{eq:W22-soln} and \eqref{eq:Wpm-soln}  exhibit several characteristic features. First, the dependence on initial conditions decays exponentially in time, reflecting relaxation toward equilibrium. Second, the relaxation rate depends explicitly on the wave number $\vq$: long‑wavelength modes equilibrate more slowly than short‑wavelength modes. Finally, at late times the correlators approach their equilibrium values universally, following power‑law behavior governed by the dimensionless combination $\vq^2\gamma\eta(\tau)\tau$. This universal late‑time scaling is discussed further in  Sec.~\ref{sec:latetime}.  

\subsection{Three-point correlators}

The structure of three‑point correlators is kinematically richer than that of two‑point functions, as they depend on three wave vectors $\vq_1,\vq_2,\vq_3$ constrained by $\vq_1+\vq_2+\vq_3=0$. The slow modes correspond to transverse modes defined with respect to each wave vector. We parameterize them in spherical coordinates (the same coordinates used in Ref. \cite{akamatsu}) as
\begin{eqnarray}
        \hat \vq_i \equiv \frac{\vq_i}{ |\vq_i|}&\equiv&(\sin \theta_i \cos\phi_i, \sin\theta \sin\phi_i, \cos\theta_i)   \\
        \hat{t}_1(\vq_i) &\equiv& (-\sin\phi_i,\cos\phi_i,0)    \\
        \hat{t}_2(\vq_i)&\equiv&(\cos\theta \cos\phi_i,\cos\theta \sin\phi_i,-\sin\theta_i)
\end{eqnarray}
To simplify our expressions we introduce a shortcut notation
\begin{equation}
   s_i\equiv \sin\theta_i^2\quad{\rm for}\,\,i=1,2,3,\quad s_{23}\equiv \sin\theta_2^2+\sin\theta_3^2,\quad s_{123}\equiv \sin\theta_1^2+\sin\theta_2^2+\sin\theta_3^2\,,
\end{equation}
and denote the effective wave-vector that determines the relaxation rate as 
\begin{equation}
    q^2\equiv \vq_1^2+\vq_2^2+\vq_3^2\,.
    \label{eq:q2}
\end{equation}

The evolution equation for transverse three‑point correlators admits a compact solution,  which naturally separates into two parts: an exponentially decaying contribution proportional to the initial condition, and a dynamical contribution. The latter is expressed as a product of two object: the kinematic part, ${\cal K}^{[abc]}(\vq_1,\vq_2,\vq_3)$, that only depends on the wave-vectors but not $\tau$, and the dynamical part, ${\cal I}^{[abc]}(\vq_1,\vq_2,\vq_3;\tau)$: 
\begin{equation}
\label{eq:W3-exact}
\W{a}{b}{c}({\bm q}_1,{\bm q}_2,{\bm q}_3;\tau)=
\frac{\W{a}{b}{c}({\bm q}_1,{\bm q}_2,{\bm q}_3;\tau_{\rm in})}{\hat\tau^{3+s}}e^{-\left(\Gamma(q^2,\hat\tau)-\Gamma(q^2,1)\right)} + 3i\left[ 
{\cal K}^{[abc]}(\vq_1,\vq_2,\vq_3){\cal I}^{[abc]}(\vq_1,\vq_2,\vq_3;\tau) 
\right]_{\overline{123}}
\end{equation}   
The permutation affects all the indices, including the wave-vectors as usual. The parameters $s$ for different combinations of the indices is given as
\begin{equation}
s =
\begin{cases}
0 & \text{for } \W{1}{1}{1}, \\
s_3 & \text{for } \W{1}{1}{2}, \\
s_{23} &\text{for } \W{1}{2}{2}, \\
s_{123} &\text{for } \W{2}{2}{2}.
\end{cases}
\end{equation}
The kinematical factors result from projecting on the plane transverse to the relevant wave-vector $\vq_i$ and are given by
\begin{eqnarray}
\label{eq:K1}
     {\cal K}^{[1bc]}(\vq_1,\vq_2,\vq_3)&\equiv& \vq_3\cdot \hat t_b(\vq_2) \hat t_a(\vq_1) \cdot \hat t_c(\vq_3)
     +\vq_2\cdot \hat t_c(\vq_3) \hat t_a(\vq_1) \cdot \hat t_b(\vq_2)
     \\
     {\cal K}^{[2bc]}(\vq_1,\vq_2,\vq_3)&\equiv&  \vq_1\cdot \hat t_b(\vq_2) \hat t_a(\vq_1) \cdot \hat t_c(\vq_3)+
     \vq_2\cdot \hat t_a(\vq_1) \hat t_b(\vq_2) \cdot \hat t_c(\vq_3) \,.
     \label{eq:K2}
\end{eqnarray}
The dynamical contributions follow from the inhomogeneous part in the evolution equation, Eq. \eqref{eq:W3-eqn}, sourced by the product of two-point functions encoding the nonlinear coupling between Gaussian and non‑Gaussian fluctuations:
\begin{eqnarray}
\label{eq:I1bc}
     {\cal I}^{[1bc]}(\vq_1,\vq_2,\vq_3;\tau)&=&\frac{e^{-\Gamma(q^2, \hat\tau)}}{\tau^{3+s}}\int_{\tau_{\rm in}}^\tau du\frac{u^{3+s}}{w(u)} \,e^{\Gamma(q^2, u/\tau_{\rm in})}
     \w{b}{b}(\vq_2;u)\w{c}{c}(\vq_3;u) \\
     {\cal I}^{[2bc]}(\vq_1,\vq_2,\vq_3;\tau)&=&\frac{ e^{-\Gamma(q^2, \hat\tau)}}{\tau^{3+s}}\int_{\tau_{\rm in}}^\tau du\frac{u^{3+s}}{w(u)} \,e^{\Gamma(q^2, u/\tau_{\rm in})}\w{2}{2}(\vq_1;u)\w{b}{b}(\vq_2;u)\,.
\label{eq:I2bc}
\end{eqnarray}
The effective relaxation rate for the three‑point functions is controlled by the combined wave-vector  $q^2=\vq_1^2+\vq_2^2+\vq_3^2$, implying that small wave-vector configurations dominate late‑time behavior.

The three‑point correlators vanish in thermal equilibrium due to isotropy. This can also be verified as follows.  At late times, the two-point functions in Eqs. \eqref{eq:I1bc} and \eqref{eq:I2bc} approach their $\vq$-independent equilibrium values.  Therefore for large $\tau$, the dynamical terms, ${\cal I}^{[abc]}(\vq_1,\vq_2,\vq_3;\tau)$, become independent of $\vq_i$.  At the same time, the kinematic part vanishes:
\begin{equation}
\label{eq:Kperm}
    \left[ {\cal K}^{[abc]}(\vq_1,\vq_2,\vq_3) \right]_{\overline{123}}=0
\end{equation} due to the constraint $\vq_1+\vq_2+\vq_3=0$. 

Although the three-point functions eventually relax to zero, there is a non-trivial transitionary stage sourced by the out-of-equilibrium dynamics of the two-point functions as shall illustrate. Furthermore, the late-time behavior is given by a universal power law which we discuss in the next section.

A notable feature of the three‑point correlators is that they are complex in momentum space. This arises from the purely imaginary structure of the nonlinear coupling coefficients in the evolution equation. After inverse Wigner transformation to position space, however, the resulting correlators are real, consistent with the reality of the underlying hydrodynamic fields.

Before we discuss the universal late-time behavior of the three-point functions, we present a numerical illustration of the evolution of three-point functions.
We initialize  all the correlators at their equilibrium values. Because of the expansion, as the system evolves they deviate from their equilibrium values at intermediate times and then relax back to equilibrium at late times. Modes with longer wavelengths equilibrate slower than the those with shorter wavelengths. The numerical solutions illustrate this behavior. 

\begin{figure}
\includegraphics[scale=0.5]{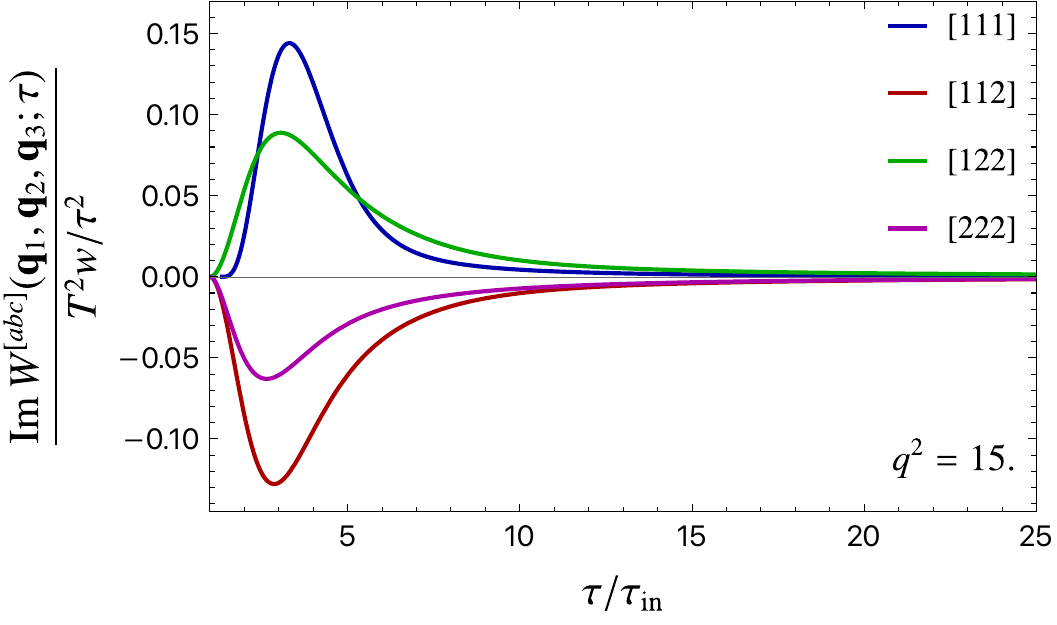}
\includegraphics[scale=0.5]{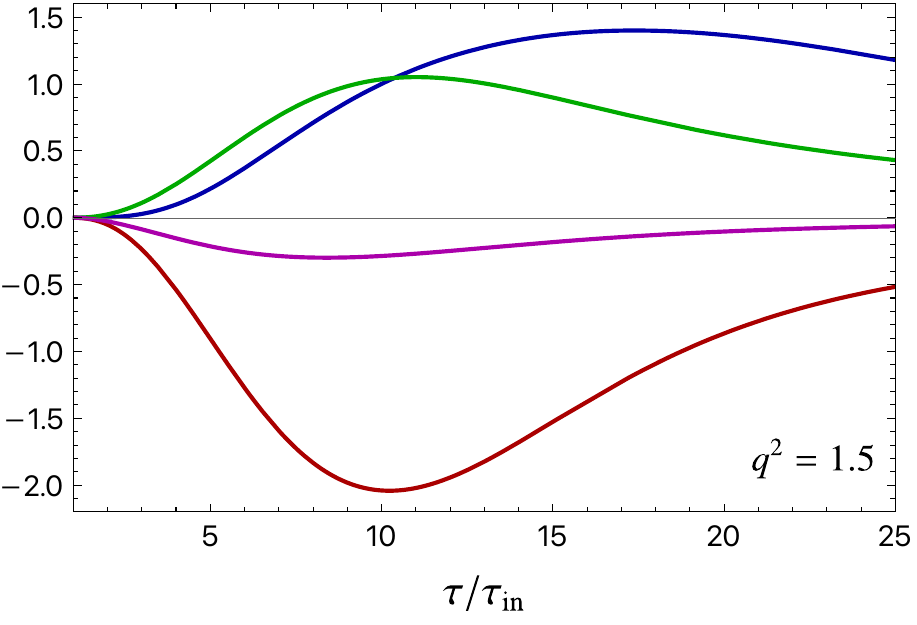}
\caption{The three-point correlator (normalized to be dimensionless) of transverse momentum fluctuations for two different values of $q^2$. The smaller $q^2$ modes relax slower and are greater in magnitude.  }
\label{fig:W3}
\end{figure}

Fig.~\ref{fig:W3} shows the time evolution of the four transverse modes whose initially are set to their equilibrium values (i.e. zero). To make the three-point functions dimensionless in the plots, we rescaled them with the factor $\left(W_2^{\rm eq}\right)^2/\tau=T^2 w/\tau^2$. This combination corresponds to the typical scale of the inhomogeneous term in the evolution equation given in Eq.~\eqref{eq:W3-eqn}. For illustrative purposes, we chose two sets of values of $\vq_i$s which correspond to $q^2=15$ and $q^2=1.5$, in arbitrary units. To represent a generic case where all $\vq_i$s are oriented in different directions, we set $\theta_1=\pi/4$, $\phi_1=0$, $\theta_2=\pi/8$, $\phi_2=\pi/3$. Finally we chose $\eta/s=1/(2\pi)$. 

\begin{figure}
\includegraphics[scale=0.5]{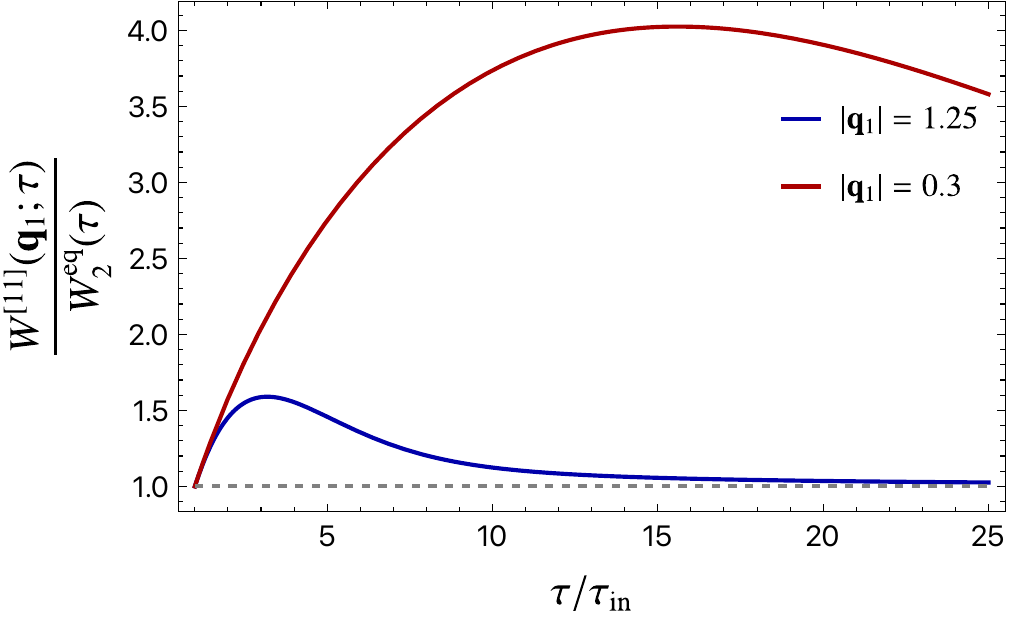}
\caption{The two-point correlator (normalized to its equilibrium value) fluctuations for two different values of $|\vq_1|^2$. The smaller $|\vq_1|^2$ modes relax slower and are greater in magnitude. }
\label{fig:W2}
\end{figure}

As mentioned above, even though the three-point correlators are initialized at their equilibrium values, as the fluid expands, they deviate from their equilibrium values. This behavior is inherited from the evolution of the two-point functions. From Eqs.~\eqref{eq:I1bc} and \eqref{eq:I2bc}, one can see that the three-point functions are sourced by the product of two $W_2$s with different $\vq_i$s. Since $W_2$s are also initialized at their equilibrium values, which are independent of $\vq_i$, their contribution to $W_3$ vanishes due to Eq. \eqref{eq:Kperm}. As the fluid expands, the equilibrium value of  $W_2$ changes and $W_2$ falls out of equilibrium. With the ongoing expansion the equilibrium value of $W_2$ keeps changing and different wavelengths catch up to the new equilibrium value at different rates. Therefore, in general, $W_2$s develop a $\vq$ dependence over time which then sources the $W_3$ via Eqs.~\eqref{eq:I1bc} and \eqref{eq:I2bc}. As an illustration, in Fig.~\ref{fig:W2} we show the evolution of $W^{[11]}$ for two different values of $\vq_1$s that are a part of our parameter set for the three-point functions shown in Fig.\ref{fig:W3}. 

\subsection{Late-time expansion}
\label{sec:latetime}
At late times, the correlators have a generic fall-off governed by the damping parameter $ q^2 \int d\tau \gamma_\eta(\tau)\sim q^2 \gamma_\eta(\tau) \tau$. 
For the two-point functions expanding the analytical expressions given in Eqs. \eqref{eq:W11-soln} and \eqref{eq:W22-soln} at late times yields
\begin{eqnarray}
\label{eq:W11-late-time}
\frac{\w{1}{1}(\vq;\tau)}{T(\tau)w(\tau)/\tau}&\sim & 1+\frac{c_s^2}{\vq^2\gamma_\eta(\tau)\tau}+\frac{(3c_s^2+1)c_s^2}{2(\vq^2\gamma_\eta(\tau)\tau)^2}+\frac{(3c_s^2+1)(2c_s^2+1)c_s^2}{2(\vq^2\gamma_\eta(\tau)\tau)^3}+\dots 
\\
\frac{\w{2}{2}(\vq;\tau)}{T(\tau)w(\tau)/\tau}&\sim & 1+\frac{c_s^2-s}{\vq^2\gamma_\eta(\tau)\tau}+\frac{(3c_s^2+1-2s)(c_s^2-s)}{2(\vq^2\gamma_\eta(\tau)\tau)^2}+\frac{(3c_s^2+1-2s)(2c_s^2+1-s)(c_s^2-s)}{2(\vq^2\gamma_\eta(\tau)\tau)^3}+\dots
\label{eq:W22-late-time}
\end{eqnarray}
here $s\equiv\sin^2\theta$ since the two-point correlators depend on a single  wave-vector, $\vq$, whose polar angle is $\theta$. These expressions agree with the earlier results \cite{akamatsu}. 

\begin{figure}[h]
\includegraphics[scale=0.6]{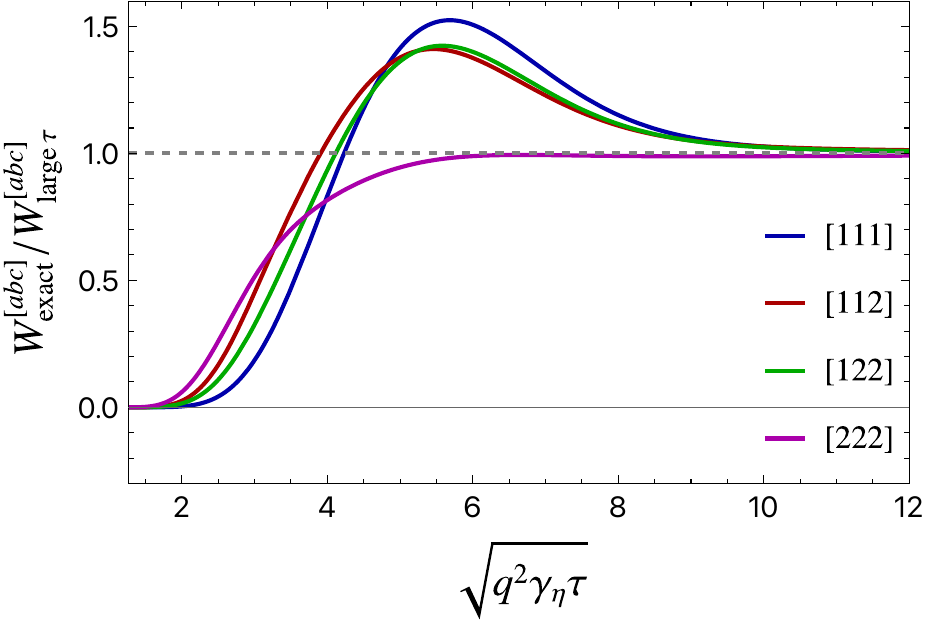}
\caption{The ratio of the three-point correlator to the late time expansion given in Eq.~\eqref{eq:W3-late-time} up to NNLO. }
\label{fig:late-time}
\end{figure}

In equilibrium, the three-point functions approach their equilibrium value (zero) in a universal way as well.  Expanding the integrals in Eqs. \eqref{eq:I1bc} and \eqref{eq:I2bc} at late times we find
\begin{equation}
    {\cal I}^{[abc]}(\vq_1,\vq_2,\vq_3;\tau)=\frac{T^2(\tau)w(\tau)}{\tau} \sum_{n=2}^\infty  \frac{c^{[abc]}_n(\vq_1,\vq_2,\vq_3)}{[q^2\gamma_\eta(\tau)\tau]^{n}}\,.
\end{equation}
The coefficients in this case are complicated combinations of the wave-vectors whose first three terms are listed in Appendix \ref{sec:cns}. The effective wave vector, $\vq$, is given in Eq. \eqref{eq:q2}.  Multiplying with the appropriate kinematic factors given in Eqs. \eqref{eq:K1} and \eqref{eq:K2} leads to the expression
\begin{equation}
    \W{a}{b}{c}(\vq_1,\vq_2,\vq_3;\tau)=i\frac{T^2(\tau)w(\tau)}{\tau} \sum_{n=2}^\infty  \frac{C^{[abc]}_n(\vq_1,\vq_2,\vq_3)}{[q^2\gamma_\eta(\tau)\tau]^{n}}
    \label{eq:W3-late-time}
\end{equation}
where
\begin{equation}
   C^{[abc]}(\vq_1,\vq_2,\vq_3)\equiv 3\left[ 
{\cal K}^{[abc]}(\vq_1,\vq_2,\vq_3)c^{[abc]}_n(\vq_1,\vq_2,\vq_3;\tau) 
\right]_{\overline{123}}
\end{equation}
In Fig. \ref{fig:late-time} we show how the late-time expansion converges to the exact result for the same set of parameters that we used to plot the three-point function in Fig. \ref{fig:W3} with $q^2=15$. Notice that we chose an effective time variable $\sqrt{q^2\geta\tau}$ which is the natural expansion parameter in Eq.~\eqref{eq:W3-late-time}. This is a three-point function analogue  that the transition to the universal late-time behavior happens at $\sqrt{q^2\geta\tau}\sim {\cal O}.(1)$. 

\section{Conclusions and Outlook}
\label{sec:conclusions}

We studied the non‑equilibrium evolution of non‑Gaussian hydrodynamic fluctuations in a relativistic fluid undergoing Bjorken expansion. Using the effective field theory formulation of fluctuating hydrodynamics, we derived deterministic evolution equations for equal‑time two‑ and three‑point correlation functions of energy and momentum density fluctuations.

A central element of our analysis is the use of the average Landau frame, which enables a consistent treatment of velocity fluctuations and avoids ill‑defined contributions from time derivatives of stochastic noise that arise in higher‑order correlators when working in fluctuating local rest frames. In the Bjorken background, this frame coincides with the density frame, allowing for a particularly transparent formulation of the evolution equations. Within this setup, we obtained the explicit 
form of the effective Lagrangian to cubic order and  its consistency with dynamical KMS symmetry.
We then derived the evolution equations for the equal time three-point hydrodynamic correlators. As a nontrivial check, we demonstrated that the equilibrium correlators, calculated independently from the entropy functional in the density frame, emerge as stationary solutions through delicate cancellations between dissipative and multiplicative noise terms, as well as kinematical factors.

Exploiting the symmetries of Bjorken flow, we analyzed the spectrum of fluctuation modes explicitly and showed that only transverse momentum three‑point modes are slow, while longitudinal modes undergo rapid oscillations and average out over expansion time scales. Focusing on the slow modes, we derived analytic solutions to the evolution equations and illustrated their behavior numerically for representative parameter choices. The three‑point correlators are dynamically generated by the out‑of‑equilibrium evolution of the two‑point functions, exhibit memory effects characteristic of expanding media, and relax to zero at late times as required by isotropy. Both Gaussian and non‑Gaussian correlators approach equilibrium following a universal power‑law behavior governed by the relaxation rate $q^2\geta(\tau)\tau$,  independent of initial conditions.

Complementary to recent work, \cite{An:2026glk}, where evolution equations for non‑Gaussian fluctuations in arbitrary backgrounds were derived by using the confluent formalism, our explicit analytical results constitute the first step towards providing 
analytically controlled input for incorporating the dynamics of non‑Gaussian fluctuations into the predictions for critical point signals via frameworks such as maximum‑entropy freeze‑out . Extensions to include conserved charges, more general expanding backgrounds, and critical dynamics constitute important directions left for future work.

\begin{acknowledgments}
This work is supported by the National Science Foundation CAREER Award PHY-2143149.
\end{acknowledgments}

\appendix

\section{Operators}
\label{sec:operators}
Below we list the differential operators that appear in the evolution equations for the two and three-point functions in position space given in Eqns. \eqref{eq:G2-evolution} and \eqref{eq:G3-evolution}.

\begin{align}
\begin{split}
&L^{\varepsilon}_{\varepsilon}(\vx)\,\delta\varepsilon(\vx)
= -\frac{1+c_s^2}{\tau}\,\delta\varepsilon(\vx)
\;\;,\;\;
L^{\varepsilon}_{i}(\vx)\,\delta\pi^i(\vx)
= -\partial_i\delta\pi^i(\vx)
\;\;,\;\;
L^{i}_{\varepsilon}(\vx)\,\delta\varepsilon(\vx)
= -c_s^2\,\partial^i\delta\varepsilon(\vx)
\\[0.5ex]
&L^{i}_{j}(\vx)\,\delta\pi^j(\vx)
= -\frac{1}{\tau}\,\delta\pi^i(\vx)
-\frac{1}{\tau}\,\delta^i_{\;3}\,\delta\pi^3(\vx)
+\gamma_\eta\,\partial^l\partial_l\delta\pi^i(\vx)
+\left(\gamma_\zeta+\frac{\gamma_\eta}{3}\right)\partial^i\partial_j\delta\pi^j(\vx)
\end{split}
\\[1.0ex]
\begin{split}
&L^{\varepsilon}_{\varepsilon\varepsilon}(\vx)\,\delta\varepsilon(\vx)\,\delta\varepsilon(\vx)
= -\frac{p^{\prime\prime}}{2\tau}\,\bigl(\delta\varepsilon(\vx)\bigr)^2
\;\;,\;\;
L^{\varepsilon}_{ij}(\vx)\,\delta\pi^i(\vx)\,\delta\pi^j(\vx)
= \frac{c_s^2}{\tau w}\,\delta\pi^l(\vx)\,\delta\pi_l(\vx)
-\frac{1}{\tau w}\delta\pi^3\delta\pi^3
\\[0.5ex]
&L^{i}_{\varepsilon\varepsilon}(\vx)\,\delta\varepsilon(\vx)\,\delta\varepsilon(\vx)
= -\frac{1}{2}\,p^{\prime\prime}\,\partial^i\!\bigl(\delta\varepsilon(\vx)\bigr)^2
\\[0.5ex]
&L^{i}_{jk}(\vx)\,\delta\pi^j(\vx)\,\delta\pi^k(\vx)
= -\frac{1}{w}\partial_l\left(\delta\pi^l(\vx)\delta\pi^i(\vx)\right)
+\frac{c_s^2}{w}\,\partial^i\!\left(\delta\pi^l(\vx)\,\delta\pi_l(\vx)\right)
\\[0.5ex]
&L^{i}_{\varepsilon j}(\vx)\,\delta\varepsilon(\vx)\,\delta\pi^j(\vx)
= -\gamma_\eta\,\left(\frac{w'}{w}+\frac{T'}{T}\right)\,\partial^l\partial_l\!\Bigl(\delta\varepsilon(\vx)\,\delta\pi^i(\vx)\Bigr)
-\left(\gamma_\zeta+\frac{\gamma_\eta}{3}\right)\left(\frac{w'}{w}+\frac{T'}{T}\right)\,
\partial^i\partial_l\!\Bigl(\delta\varepsilon(\vx)\,\delta\pi^l(\vx)\Bigr)
\\
&\qquad\qquad
+\frac{1}{Tw}\,(\eta T)'\,\partial^l\!\Bigl(\delta\varepsilon(\vx)\,\partial_l\delta\pi^i(\vx)\Bigr)
+\frac{1}{Tw}\,(\eta T)'\,\partial_l\!\Bigl(\delta\varepsilon(\vx)\,\partial^i\delta\pi^l(\vx)\Bigr)
+\frac{1}{Tw}\,\Bigl(\bigl(\zeta-\tfrac{2}{3}\eta\bigr)T\Bigr)'\,
\partial^i\!\Bigl(\delta\varepsilon(\vx)\,\partial_l\delta\pi^l(\vx)\Bigr)
\end{split}
\\
\begin{split}
 & Q^{ij}(\vx_1,\vx_2) = -T\eta\partial_{l_1}\partial^{l_1}\delta^{ij}(\vx_2-\vx_1) - T(\zeta+\frac{1}{3}\eta)\partial^{i_1}\partial_{l_1}\delta^{lj}(\vx_2-\vx_1) 
 \\
& Q^{ij}_{\varepsilon}(\vx_1,\vx_2)\delta\varepsilon(\vx_1) = - (T\eta)'\partial^{l_1}(\delta\varepsilon(\vx_1)\partial_{l_1}\delta^{ij}(\vx_2-\vx_1)) - (T\eta)'\partial_{l_1}(\delta\varepsilon(\vx_1)\partial^{i_1}\delta^{lj}(\vx_2-\vx_1))  \\
&\qquad\qquad
 - (T(\zeta-\frac{2}{3}\eta))'\partial^{i_1}(\delta\varepsilon(\vx_1)\partial_{l_1}\delta^{lj}(\vx_2-\vx_1))
\end{split}
\end{align}

The derivative $\partial_{l1}$ indicates that it only acts on variables with spatial coordinate $x_1$.

\section{Late-time expansion coefficients}
\label{sec:cns}
In this appendix we list the first three terms in the late-time expansion of the three-point functions given in Eq.~\eqref{eq:W3-late-time}. To optimize our we introduce the rescaled wave-vectors: 
\begin{equation}
\tilde \vq_i\equiv \frac{\vq_i}{q}=\frac{\vq_i}{\sqrt{\vq_1^2+\vq_2^2+\vq_3^2}} 
\end{equation}
The coefficients are then given by 
\begin{eqnarray}
   c^{[111]}_2(\vq_1,\vq_2,\vq_3)&=& c_s^2\left(\frac{1}{ \tvq_2^2}+\frac{1}{ \tvq_3^2}\right)
   \\
   c^{[111]}_3(\vq_1,\vq_2,\vq_3)&=& 
   \frac{1}{2}(3c_s^2+1)c_s^2\left(\frac{1}{\tvq_2^4} 
   +\frac{1}{\tvq_3^4}\right)
   +\frac{c_s^4}{\tvq_2^2\tvq_3^2}+(5c_s^2+1)c_s^2\left(\frac{1}{ \tvq_2^2}+\frac{1}{ \tvq_3^2}\right)
   \\ 
   c^{[111]}_4(\vq_1,\vq_2,\vq_3)&=& \frac{1}{2}(3c_s^2+1)(2c_s^2+1)c_s^2\left(\frac{1}{\tvq_2^6} +\frac{1}{\tvq_3^6}\right)
   +\frac{1}{2}(3c_s^2+1)c_s^4\left(\frac{1}{\tvq_3^2\tvq_2^4} +\frac{1}{\tvq_2^2\tvq_3^4}\right)
  \nonumber \\ &
  & +(3c_s^2+1)^2c_s^2\left(\frac{1}{\tvq_2^4} +\frac{1}{\tvq_3^4}\right)
   +\frac{2(3c_s^2+1)c_s^4}{\tvq_2^2\tvq_3^2}
   +2(3c_s^2+1)(5c_s^2+1)c_s^2\left(\frac{1}{ \tvq_2^2}+\frac{1}{ \tvq_3^2}\right) 
   \\
    c^{[211]}_2(\vq_1,\vq_2,\vq_3)&=& c_s^2\left(\frac{1}{ \tvq_1^2}+\frac{1}{ \tvq_2^2}\right)
   \\
   c^{[211]}_3(\vq_1,\vq_2,\vq_3)&=& 
   \frac{1}{2}(3c_s^2+1)c_s^2\left(\frac{1}{\tvq_1^4} 
   +\frac{1}{\tvq_2^4}\right)+\frac{c_s^4}{\tvq_1^2\tvq_2^2}
   +(5c_s^2+1-s_3)c_s^2\left(\frac{1}{ \tvq_1^2}+\frac{1}{ \tvq_2^2}\right)
   \\
   c^{[211]}_4(\vq_1,\vq_2,\vq_3) &=& \frac{1}{2}(3c_s^2+1)(2c_s^2+1)c_s^2\left(\frac{1}{\tvq_1^6} 
   +\frac{1}{\tvq_2^6}\right)
   +\frac{1}{2}(3c_s^2+1)c_s^4\left(\frac{1}{\tvq_2^2\tvq_1^4} 
   +\frac{1}{\tvq_1^2\tvq_2^4}\right)
   \nonumber\\&& 
   +\frac{1}{2}(6c_s^2+2-s_3)(3c_s^2+1)c_s^2\left(\frac{1}{\tvq_1^4} 
   +\frac{1}{\tvq_2^4}\right)
    +\frac{(6c_s^2+2-s_3)c_s^4}{\tvq_1^2\tvq_2^2}
    \nonumber\\& &
    +(6c_s^2+2-s_3)(5c_s^2+1-s_3)c_s^2\left(\frac{1}{ \tvq_1^2}+\frac{1}{ \tvq_2^2}\right)
  \\
  c^{[112]}_2(\vq_1,\vq_2,\vq_3)&=& \left(\frac{c_s^2}{ \tvq_2^2}+\frac{c_s^2-s_3}{ \tvq_3^2}\right)
   \\
   c^{[112]}_3(\vq_1,\vq_2,\vq_3)&=& 
   \frac{1}{2}\left(\frac{(3c_s^2+1)c_s^2}{\tvq_2^4} 
   +\frac{(3c_s^2+1-2s_3)(c_s^2-s_3)}{\tvq_3^4}\right)+\frac{c_s^2(c_s^2-s_3)}{\tvq_2^2\tvq_3^2}
   +(5c_s^2+1-s_3)\left(\frac{c_s^2}{ \tvq_2^2}+\frac{c_s^2-s_3}{ \tvq_3^2}\right)\qquad
   \\
   c^{[112]}_4(\vq_1,\vq_2,\vq_3) &=& 
   \frac{1}{2}\left(\frac{(3c_s^2+1)(2c_s^2+1)c_s^2}{\tvq_2^6} 
   +\frac{(3c_s^2+1-2s_3)(2c_s^2+1-s_3)(c_s^2-s_3)}{\tvq_3^6}\right)
   \nonumber\\&&
   +\left(\frac{(3c_s^2+1)c_s^2(c_s^2-s_3)}{\tvq_3^2\tvq_1^4} 
   +\frac{(3c_s^2+1-2s_3)(c_s^2-s_3)c_s^2}{\tvq_2^2\tvq_3^4}\right) 
   \nonumber\\& &
   +\frac{1}{2}(6c_s^2+2-s_3)\left(\frac{(3c_s^2+1)c_s^2}{\tvq_2^4} 
   +\frac{(3c_s^2+1-2s_3)(c_s^2-s_3)}{\tvq_3^4}\right)
    +\frac{(6c_s^2+2-s_3)c_s^2(c_s^2-s_3)}{\tvq_2^2\tvq_3^2}
  \nonumber\\& &
  +(6c_s^2+2-s_3)(5c_s^2+1-s_3)\left(\frac{c_s^2}{ \tvq_2^2}+\frac{c_s^2-s_3}{ \tvq_3^2}\right)
  \end{eqnarray}
\begin{eqnarray}
c^{[221]}_2(\vq_1,\vq_2,\vq_3)&=& 
    \left(\frac{c_s^2}{  \tvq_1^2}+\frac{c_s^2-s_2}{ \tvq_2^2}\right)
\\
c^{[221]}_3(\vq_1,\vq_2,\vq_3)&=& 
   \frac{1}{2}\left(\frac{(3c_s^2+1)c_s^2}{\tvq_1^4} 
   +\frac{(3c_s^2+1-2s_2)(c_s^2-s_2)}{\tvq_2^4}\right)+\frac{c_s^2(c_s^2-s_2)}{\tvq_1^2\tvq_2^2}
   +(5c_s^2+1-s_{23})\left(\frac{c_s^2}{ \tvq_1^2}+\frac{c_s^2-s_2}{ \tvq_2^2}\right)\,\,\qquad
   \\
c^{[221]}_4(\vq_1,\vq_2,\vq_3) &=& 
   \frac{1}{2}\left(\frac{(3c_s^2+1)(2c_s^2+1)c_s^2}{\tvq_1^6} 
   +\frac{(3c_s^2+1-2s_2)(2c_s^2+1-s_2)(c_s^2-s_2)}{\tvq_2^6}\right)
  \nonumber \\ && 
   +\left(\frac{(3c_s^2+1)c_s^2(c_s^2-s_2)}{\tvq_2^2\tvq_1^4} 
   +\frac{(3c_s^2+1-2s_2)(c_s^2-s_2)c_s^2}{\tvq_1^2\tvq_2^4}\right) 
   \nonumber\\& &
   +\frac{1}{2}(6c_s^2+2-s_{23})\left(\frac{(3c_s^2+1)c_s^2}{\tvq_1^4} 
   +\frac{(3c_s^2+1-2s_2)(c_s^2-s_2)}{\tvq_2^4}\right)
    +\frac{(6c_s^2+2-s_{23})c_s^2(c_s^2-s_2)}{\tvq_1^2\tvq_2^2}
   \nonumber\\ & &
  +(6c_s^2+2-s_{23})(5c_s^2+1-s_{23})\left(\frac{c_s^2}{ \tvq_1^2}+\frac{c_s^2-s_2}{ \tvq_2^2}\right)
  \\
  c^{[122]}_2(\vq_1,\vq_2,\vq_3)&=& 
    \left(\frac{c_s^2-s_2}{  \tvq_2^2}+\frac{c_s^2-s_3}{ \tvq_3^2}\right)
\\
c^{[122]}_3(\vq_1,\vq_2,\vq_3)&=& 
   \frac{1}{2}\left(\frac{(3c_s^2+1-2s_2)(c_s^2-s_2)}{\tvq_2^4} 
   +\frac{(3c_s^2+1-2s_3)(c_s^2-s_3)}{\tvq_3^4}\right)+\frac{(c_s^2-s_2)(c_s^2-s_3)}{\tvq_2^2\tvq_3^2}
   \nonumber\\&&
   +(5c_s^2+1-s_{23})\left(\frac{c_s^2-s_2}{ \tvq_2^2}+\frac{c_s^2-s_3}{ \tvq_3^2}\right)
   \\
c^{[122]}_4(\vq_1,\vq_2,\vq_3) &=& 
   \frac{1}{2}\left(\frac{(3c_s^2+1-2s_2)(2c_s^2+1-s_2)(c_s^2-s_2)}{\tvq_2^6} 
   +\frac{(3c_s^2+1-2s_3)(2c_s^2+1-s_3)(c_s^2-s_3)}{\tvq_3^6}\right)
   \nonumber\\ & &
   +\left(\frac{(3c_s^2+1-2s_2)(c_s^2-s_2)(c_s^2-s_3)}{\tvq_3^2\tvq_2^4} 
   +\frac{(3c_s^2+1-2s_3)(c_s^2-s_3)(c_s^2-s_2)}{\tvq_2^2\tvq_3^4}\right) 
   \nonumber\\& &
   +\frac{1}{2}(6c_s^2+2-s_{23})\left(\frac{(3c_s^2+1-2s_2)(c_s^2-s_2)}{\tvq_2^4} 
   +\frac{(3c_s^2+1-2s_3)(c_s^2-s_3)}{\tvq_3^4}\right)
   \nonumber\\ && 
    +\frac{(6c_s^2+2-s_{23})(c_s^2-s_2)(c_s^2-s_3)}{\tvq_2^2\tvq_3^2}
   \nonumber\\ & &
  +(6c_s^2+2-s_{23})(5c_s^2+1-s_{23})\left(\frac{c_s^2-s_2}{ \tvq_2^2}+\frac{c_s^2-s_3}{ \tvq_3^2}\right)
\\
  c^{[222]}_2(\vq_1,\vq_2,\vq_3)&=& 
    \left(\frac{c_s^2-s_1}{  \tvq_1^2}+\frac{c_s^2-s_2}{ \tvq_2^2}\right)
\\
c^{[222]}_3(\vq_1,\vq_2,\vq_3)&=& 
   \frac{1}{2}\left(\frac{(3c_s^2+1-2s_1)(c_s^2-s_1)}{\tvq_1^4} 
   +\frac{(3c_s^2+1-2s_2)(c_s^2-s_2)}{\tvq_2^4}\right)+\frac{(c_s^2-s_1)(c_s^2-s_2)}{\tvq_1^2\tvq_2^2}
  \nonumber \\&&
   +(5c_s^2+1-s_{123})\left(\frac{c_s^2-s_1}{ \tvq_1^2}+\frac{c_s^2-s_2}{ \tvq_2^2}\right)
  \\
c^{[222]}_4(\vq_1,\vq_2,\vq_3) &=& 
   \frac{1}{2}\left(\frac{(3c_s^2+1-2s_1)(2c_s^2+1-s_1)(c_s^2-s_1)}{\tvq_1^6} 
   +\frac{(3c_s^2+1-2s_2)(2c_s^2+1-s_2)(c_s^2-s_2)}{\tvq_2^6}\right)
   \nonumber\\ & &
   +\left(\frac{(3c_s^2+1-2s_1)(c_s^2-s_1)(c_s^2-s_2)}{\tvq_2^2\tvq_1^4} 
   +\frac{(3c_s^2+1-2s_2)(c_s^2-s_2)(c_s^2-s_1)}{\tvq_1^2\tvq_2^4}\right) 
  \nonumber \\& &
   +\frac{1}{2}(6c_s^2+2-s_{123})\left(\frac{(3c_s^2+1-2s_1)(c_s^2-s_1)}{\tvq_1^4} 
   +\frac{(3c_s^2+1-2s_2)(c_s^2-s_2)}{\tvq_2^4}\right)
  \nonumber \\ & &
    +\frac{(6c_s^2+2-s_{123})(c_s^2-s_1)(c_s^2-s_2)}{\tvq_1^2\tvq_2^2}
  \nonumber \\ & &
  +(6c_s^2+2-s_{123})(5c_s^2+1-s_{123})\left(\frac{c_s^2-s_1}{ \tvq_1^2}+\frac{c_s^2-s_2}{ \tvq_2^2}\right)
\end{eqnarray}

\section{Stochastic Hamiltonian Dynamics}
\label{sec:brownian}

In this appendix we illustrate the derivation of the  deterministic evolution equations from a path integral perspective in a simple Brownian motion example. Consider a particle with a Hamiltonian
\begin{equation}
    H = \frac{p^2}{2m} + V(q)
\end{equation}
undergoing Brownian motion. Let $\xi$ denote the state of the particle in phase space:
\begin{equation}
    \xi^i\equiv(q^i,p^i)
\end{equation}
The equation of motion can be compactly expressed as 
\begin{equation}
    \dot{\xi}^i = \varepsilon_{ij}\frac{\partial H}{\partial\xi^j}
\end{equation}
with 
\begin{equation}
    \varepsilon_{ij}\equiv
    \begin{pmatrix}
        0 &I_3 \\
        -I_3 &0
    \end{pmatrix}
\end{equation}
Let us now introduce a dissipative force of the form
\begin{equation}
    F_D^i = -D^{ij}\frac{\partial H}{\partial\xi^j}
\end{equation}
As a specific case, for example, the familiar drag force can be written as
\begin{equation}
    F^i_{drag} = -Dv^i = -\frac{D}{m}p^i 
\end{equation}
where the matrix $D_{ij}$ in this case reduces to
\begin{equation}
    D^{ij} = D
    \begin{pmatrix}
        0 &0 \\
        0 &I_3
    \end{pmatrix}
\end{equation}
The Langevin equation that describes the Brownian motion is given by
\begin{equation}
    \dot{\xi^i} = (\varepsilon_{ij} - D_{ij})\frac{\partial H}{\partial\xi^j} + \eta^i \equiv F^i+\eta^i
    \label{eq:Langevin}
\end{equation}
where we have introduced the noise $\eta^i$ which satisfies
\begin{equation}
    \ang{\eta_t^i\eta_{t'}^j} = 2TD^{ij}\delta(t-t')
\end{equation}
whose magnitude is fixed by the fluctuation-dissipation theorem. Eq.~\eqref{eq:Langevin} has ambiguities regarding how the noise is implemented which can be resolved by choosing a particular discretization prescription \cite{Arnold:1999uza,Arnold:1999va}. The choice of the prescription is arbitrary as long as one has the correct equilibrium distribution associated with the physical problem that is being studied. With the the It\^{o} convention the discretized Langevin equation takes the form
\begin{equation}
    \xi_{t+1}^i = \xi_t^i + \Delta tF^i + \sqrt{\Delta t}\eta^i ,\qquad F^i=(\varepsilon^{ij}-D^{ij})\frac{\partial H}{\partial\xi^j}
    \label{eq:brownian-langevin}
\end{equation}
with the following noise correlator:
\begin{equation}
    \ang{\eta_t^i\eta_{t'}^j} = 2TD^{ij}\delta_{tt'}
\end{equation}

One can also study the stochastic system via the equations of motions for the moments of the underlying probability distribution. Let $P(t,\xi)$ the probability distribution in the phase space. Its time evolution is described by the the Fokker-Planck equation
\begin{equation}
    \partial_t P + \{P,H \} =\frac{\partial}{\partial\xi^j} \Big[ PD^{ij}\frac{\partial H}{\partial\xi^j} + \frac{\partial}{\partial\xi^j}(TD^{ij}P) \Big]\,.
\end{equation}
which admits the equilibrium distribution,
\begin{equation}
    P_{\rm eq} = e^{-H/T},
\end{equation}
as a stationary solution. From the Fokker-Planck equation, it is possible write down the time evolution equations for the expectation value of an arbitrary function 
\begin{equation}
    \ang{f}_t = \int_\xi P(t,\xi)f(\xi)
\end{equation}
For example, the Brownian motion analogs of the equal time two-point functions satisfy
\begin{eqnarray}
\label{eq:brownian-2pt-1}
        \partial_t\ang{q_iq_j} &=& \frac{1}{m}\ang{p_iq_j + p_jq_i} \\
\label{eq:brownian-2pt-2}
        \partial_t\ang{q_ip_j} &=& \ang{\frac{1}{m}p_ip_j - q_iV_{,j} - \frac{1}{m}q_iD_{jk}p^k} \\
\label{eq:brownian-2pt-3}
        \partial_t\ang{p_ip_j} &=& -\ang{V_{,i}p_j + p_iV_{,j}+\frac{1}{m}\left(D_{ik}p^kp_j + p_iD_{jk}p^k \right) } + 2TD_{ij}
\end{eqnarray}
where $A_{,j}\equiv\partial A/\partial\xi^j$. The stationary solutions of the above equations correspond to the equilibrium correlators:
\begin{equation}
    \ang{p_iq_j}_{\rm eq}=\ang{p_iV_{,j}}_{\rm eq}0 ,\qquad  \ang{q_iV_{,j}}_{\rm eq}=\frac{1}{m}\ang{p_ip_j}_{\rm eq}=T\delta_{ij} \,.
\end{equation}
In particular, from the second equation above we can reproduce the Virial theorem
\begin{equation}
        \ang{{\bm q}\cdot{\bm \nabla}V}= d T = 2\times \frac{d}{2}T=2\ang{K}
\end{equation}
where $d=\delta^i_i$ is the number of spatial dimensions. 

The Langevin equation given in Eq.\eqref{eq:brownian-langevin} admits a path integral representation which can be generated by the partition function
\begin{equation}
        Z = \int DX_aD\xi e^{iS[\xi,X_a]}\,.
\end{equation}
where
\begin{equation}
    S=\int dt \left[ X_{ai}\left(\dot{\xi}^i-Q^{ij}\frac{\partial H}{\partial\xi^j}\right) + iTX_{ai}D^{ij}X_{aj} \right]
\end{equation}
This expression directly follows from the Martin-Siggia-Rose form, or more systematically from a Schwinger-Keldysh representation of the Brownian motion. Expressing the quadratic $X_a^iX_a^j$ term via a Hubbard-Stratonovich field and integrating out the $X_a$ field leads to Eq.\eqref{eq:brownian-langevin} as follows 
\begin{equation}
    \begin{split}
        Z &= \int DX_aD\eta D\xi e^{\int dt\{iX_{ai}(\dot{\xi}^i - Q^{ij}\frac{\partial H}{\partial\xi^j} - \eta^i) - \frac{1}{4T}\eta^i D_{ij}^{-1}\eta^j\}} \\
        &=\int D\eta D\xi e^{-\frac{1}{4T}\int dt\eta^i D_{ij}^{-1}\eta^j}\delta(\dot{\xi}^i - Q^{ij}\frac{\partial H}{\partial\xi^j}-\eta^i)
    \end{split}
\end{equation}
where the Hubbard-Stratonovich field $\eta$ satisfies
\begin{equation}
    \ang{\eta_t^i\eta_{t'}^j} = 2TD^{ij}\delta(t-t')\,.  
\end{equation}
We have also introduced $Q_{ij}\equiv \varepsilon_{ij} - D_{ij}$ for convenience.

To obtain the evolution equations for the two-point functions, we write the following Schwinger-Dyson equation 
\begin{equation}
\label{eq:brownian-SD}
    \ang{\frac{\delta S}{\delta X_{ai}(t_1)}\xi^j(t_2)}=0 \Rightarrow \ang{\dot{\xi}^i(t_1)\xi^j(t_2)} = \ang{Q^{il}\frac{\partial H}{\partial\xi^l}(t_1)\xi^j(t_2)} - 2iTD^{il}\ang{X_{al}(t_1)\xi^j(t_2)}\,.
\end{equation}
The last term contains $\ang{X_{al}\xi^j}$, which is the correlation function of r-field and a-field in the language of effective field theory ($G_{ar}$). To calculate this, first note that $X_a$ plays the role of  the ``conjugate momentum" of $\xi$:
\begin{equation}
    \pi_i=\frac{\partial H}{\partial\dot{\xi}^i} = X_{ai}
\end{equation}
which allows us to relate the $G_{ar}$ term with the canonical commutation relation,
\begin{equation} 
    \ang{X_{ai}(t_1)\xi^j(t_2)} = \ang{\pi_i(t_1)\xi^j(t_2)} = \Theta(t_2-t_1)\ang{[\xi^j(t_2),\pi_i(t_1)]} = i\Theta(t_2-t_1)\delta_i^j
\end{equation}
 The appearance of the Heavyside function indicates that the noise only affects subsequent fields.
The equal-time limit has to be taken carefully. We take the symmetric equal-time limit which amounts to averaging over $t_1-t_2\rightarrow 0^{\pm}$. This choice is equivalent to resolving the discretization ambiguity in the Langevin equation in a way that reproduces the correct equilibrium distribution. As a result,
\begin{eqnarray}
     \partial_t\ang{\xi^i(t)\xi^j(t)} = \frac{1}{2}\lim_{\Delta t\rightarrow0} &&\left( 
\ang{Q^{il}\frac{\partial H}{\partial\xi^l}(t+\Delta t)\xi^j(t)} - 2iTD^{il}\ang{X_{al}(t+\Delta t)\xi^j(t)}\right.
\nonumber \\ &&
\left.
\quad+\ang{Q^{il}\frac{\partial H}{\partial\xi^l}(t)\xi^j(t+\Delta t)} - 2iTD^{il}\ang{X_{al}(t)\xi^j(t+\Delta t)}\,+(i\rightarrow j)
     \right)
\end{eqnarray}
Here the second term in the second row vanishes due to the Heavyside function and we obtain 
\begin{equation}
\label{eq:brownian-sd-equaltime}
    \partial_\tau\ang{\xi^i\xi^j} = \ang{Q^{il}\frac{\partial H}{\partial\xi^l}\xi^j} + \ang{\xi^iQ^{jl}\frac{\partial H}{\partial\xi^l}} + 2TD^{ij}
\end{equation}
Expressing Eq.~\eqref{eq:brownian-sd-equaltime} in $q^i$ and $p^i$ reproduces the evolution equations \eqref{eq:brownian-2pt-1}-\eqref{eq:brownian-2pt-3} we derived earlier.

\bibliography{references}

\end{document}